\documentclass[%
reprint,
amsmath,amssymb,
aps,
]{revtex4-2}

\usepackage{natbib}

\usepackage{graphicx}%
\usepackage{dcolumn}%
\usepackage{bm}%

\usepackage{pdfsync}		%
\usepackage{amsmath}		%
\usepackage{amssymb}		%
\usepackage{bm}             %
\usepackage{graphicx}		%
\usepackage[footnotesize]{caption}
\usepackage[listofformat=subsimple,
labelformat=simple]{subfig}
\usepackage{psfrag}   	%
\usepackage{color}			%
\usepackage{balance}    %
\usepackage{mathnotation} %
\usepackage{booktabs}
\usepackage{multirow}
\usepackage{mathtools}

\usepackage{marginnote}

\makeatletter
\renewcommand*\env@matrix[1][*\c@MaxMatrixCols c]{%
\hskip -\arraycolsep
\let\@ifnextchar\new@ifnextchar
\array{#1}}
\makeatother

\begin{document}

\preprint{APS/123-QED}

\title{Optimal Gaits for Drag-dominated Swimmers with Passive Elastic Joints}%

\author{Suresh Ramasamy}
\email{ramasams@oregonstate.edu}%
\author{Ross L. Hatton}%
\email{hattonr@oregonstate.edu}
\affiliation{%
Collaborative Robotics and Intelligent Systems (CoRIS) Institute,\\
Oregon State University, Corvallis, OR, USA
}%

\date{\today}%

\begin{abstract}
In this paper, we identify optimal swimming strategies for drag-dominated swimmers with a passive elastic joint. We use resistive force theory (RFT) to obtain the dynamics of the system. We then use frequency domain analysis to relate the motion of the passive joint to the motion of the actuated joint. We couple this analysis with elements of the geometric framework introduced in our previous work aimed at identifying useful gaits for systems in drag dominated environments, to identify speed-maximizing and efficiency-maximizing gaits for drag-dominated swimmers with a passive elastic joint.
\end{abstract}

\maketitle

\section{Introduction}

\begin{figure*}%
\centering
\includegraphics[width=\textwidth]{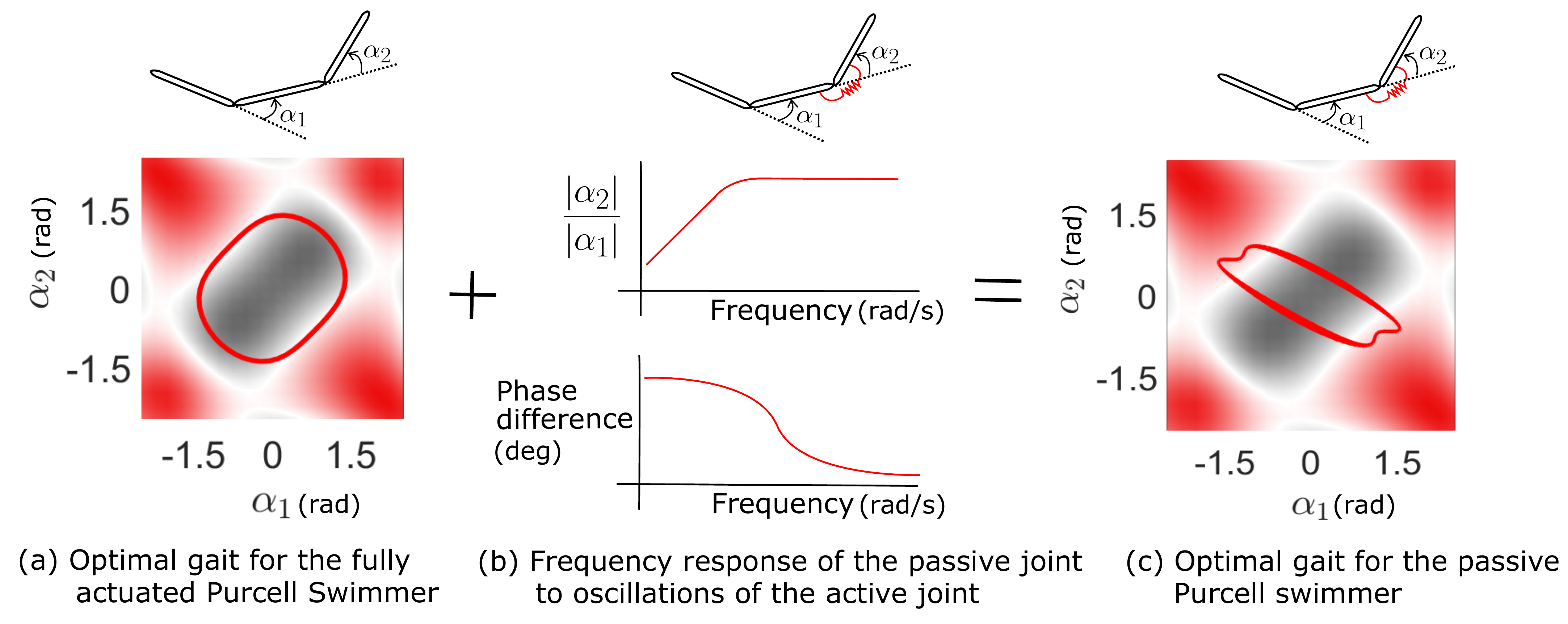}
\caption{Effect of a passive elastic joint on the shape and pacing of the (speed-maximizing) optimal gait for a Purcell swimmer. (a) The (speed and efficiency maximizing) optimal gait for the fully actuated Purcell swimmer overlaid on the curvature of the system dynamics. (b) Illustration of the Bode plot of the response of the passive elastic joint to oscillations of the active joint. This response dictates the locus of achievable gaits in the shape space for the passive swimmers. (c) The (speed maximizing) optimal gait for the passive Purcell swimmer. Whereas the optimal gait for the fully actuated Purcell swimmer is plotted with a line of uniform thickness indicating constant power dissipation throughout the cycle, the optimal gait for the passive Purcell swimmer in (c) is plotted using a line of varying thickness, with thickness at a point corresponding to the magnitude of power required at that point of the gait.}
\label{fig:coverfig}
\end{figure*}

A common strategy for locomotion among animals and robots is to couple cyclical shape changes (gaits) to an interaction with the environment. A long term research focus of the geometric mechanics community has been finding geometric principles that describe what makes a gait effective. Many geometric tools resulting from this line of research have only been developed for systems with a fully actuated shape space. Passive elastic joints, such as flexible fins and tails, however, play an important role in locomotion in many systems~\cite{brennen1977fluid}. In this paper, we expand on the geometric framework that was developed in~\cite{ramasamy2016soap,ramasamy2017geometric} to identify optimal gaits for fully actuated swimmers in low Reynolds number fluids, extending it to swimmers with passive elastic shape elements. We then use this extended framework to identify optimal gaits for swimmers in drag dominated environments with passive elastic shape elements.

Our analysis draws on a rich history of work in the geometric mechanics community aimed at using concepts from differential geometry to understand how systems locomote in a low Reynolds number fluid, what the main factors that determine the efficiency of gaits are, and what optimal gaits look like for these systems. In particular, we know that the efficiency of a gait (speed at a given power level or power required at a given speed) depends on its path, period, and pacing in the system's shape space:
\begin{packed_enum}
\ie The net displacement per cycle corresponds to the amount of ``constraint curvature" the gait encompasses. Gaits that maximize displacement per cycle thus enclose sign-definite regions of the system's constraint curvature functions (CCFs), which can be obtained from the dynamics of the system~\cite{Hatton:2013TRO:Swimming,Hatton:2015EPJ}.
\ie At a given level of average power consumption, the minimum time taken to execute one cycle of the gait corresponds to the path length of the gait trajectory under a Riemmanian metric on the shape space, meaning that shorter gaits can be executed more frequently or at a lower power level than longer gaits~\cite{TRO-Cartography}.
\ie Gaits that progress through their gait kinematics at a steady pace have a lower average power cost for a given frequency than those that ``surge'' and ``dwell'' at points in the cycle, or equivalently, can be executed at a higher frequency for a given average power cost~\cite{Becker:2003}.
\end{packed_enum}
The process of finding efficient gaits thus involves striking a balance between maximizing the enclosed constraint curvature and minimizing the metric-weighted perimeter of the gait and then finding a steady-pace parametrization of the resulting curve. In~\cite{ramasamy2016soap,ramasamy2017geometric,ramasamy2019geometry}, we built a variational framework based on this geometric insight that identifies optimal gaits for fully-actuated drag-dominated kinematic systems.

In this paper, we extend this framework to identify optimal gaits for swimmers with passive elastic joints and demonstrate the framework on systems with a single active and a single passive joint. The ways in which the dynamics of swimmers with passive elastic joints differ from the dynamics of fully actuated swimmers are:
\begin{packed_enum}
\ie Because of the coupling between the actuated and passive joints, the passive swimmers can execute only a subset of the gait kinematics the fully actuated swimmers can execute.
\ie This coupling between the actuated and passive joints also endows each gait achievable by the passive swimmer with a unique pacing. Hence, the pacing cost cannot be minimized separately from the kinematic cost.
\end{packed_enum}

As illustrated in Fig.~\ref{fig:coverfig}, we address the two problems introduced by the presence of passive elastic joints: we first use frequency domain analysis to analytically approximate the motion of the unactuated joint in response to the motion of the actuated joint. We then combine this frequency-space analysis with elements of the geometric framework introduced in~\cite{ramasamy2019geometry} to construct a gradient-descent algorithm that identifies optimal gaits and the pacing associated with these gaits for passive swimmers. The optimal gaits for passive swimmers maximize the CCF integral relative to perimeter and pacing costs, subject to amplitude and phase constraints of a first order system.

Drag-dominated swimmers with passive elements were previously studied in~\cite{passov2012dynamics,krishnamurthy2017schistosoma,montino2015three,jo2016passive,Burton:2010}. Of these works,~\cite{passov2012dynamics,krishnamurthy2017schistosoma} and~\cite{montino2015three} are most relevant to this paper, as they discuss the motion of swimmers with a harmonically-driven active joint and a passive joint. The analysis in~\cite{passov2012dynamics} is particularly relevant, where using perturbation expansion, explicit expressions for leading order solutions were derived for harmonic input oscillations, and the optimal swimmer geometry was obtained for the Purcell swimmer. In this paper, unlike in~\cite{passov2012dynamics}, we do not restrict our input to simple harmonic oscillations, and use a higher order representation of the system dynamics.

The rest of the paper is organized as follows: In~\S\ref{sec:background}, we review the geometric locomotion model and derive the dynamics of the drag-dominated swimmers. In~\S\ref{sec:frequency}, we present our frequency domain analysis to relate the motion of the passive joint to the motion of the active joint. In~\S\ref{sec:optimalgaits}, we recall a few key elements of the gradient descent algorithm introduced in~\cite{ramasamy2019geometry}, and combine them with the frequency domain analysis of~\S\ref{sec:frequency}, to set up the stroke optimization problem. In~\S\ref{sec:Optimizer}, we describe the process of finding the speed-maximizing gaits for passive swimmers. In~\S\ref{sec:energyopt}, we describe the process of finding the efficiency-maximizing gaits for the passive swimmers.

\section{System Dynamics}
\label{sec:background}
In this section, we review the dynamics of swimmers in a low Reynolds number fluid. We focus here on the geometric structure of the system dynamics, and not on the explicit expressions. A more detailed discussion of how the dynamics are obtained is presented in~\cite{Hatton:2013TRO:Swimming} and~\cite{TRO-Cartography}. Some of the the materials presented in this section also appear in~\cite{Hatton:2013TRO:Swimming} and~\cite{TRO-Cartography}.

\subsection{System geometries}
\begin{figure}%
\centering
\includegraphics[width=.5\textwidth]{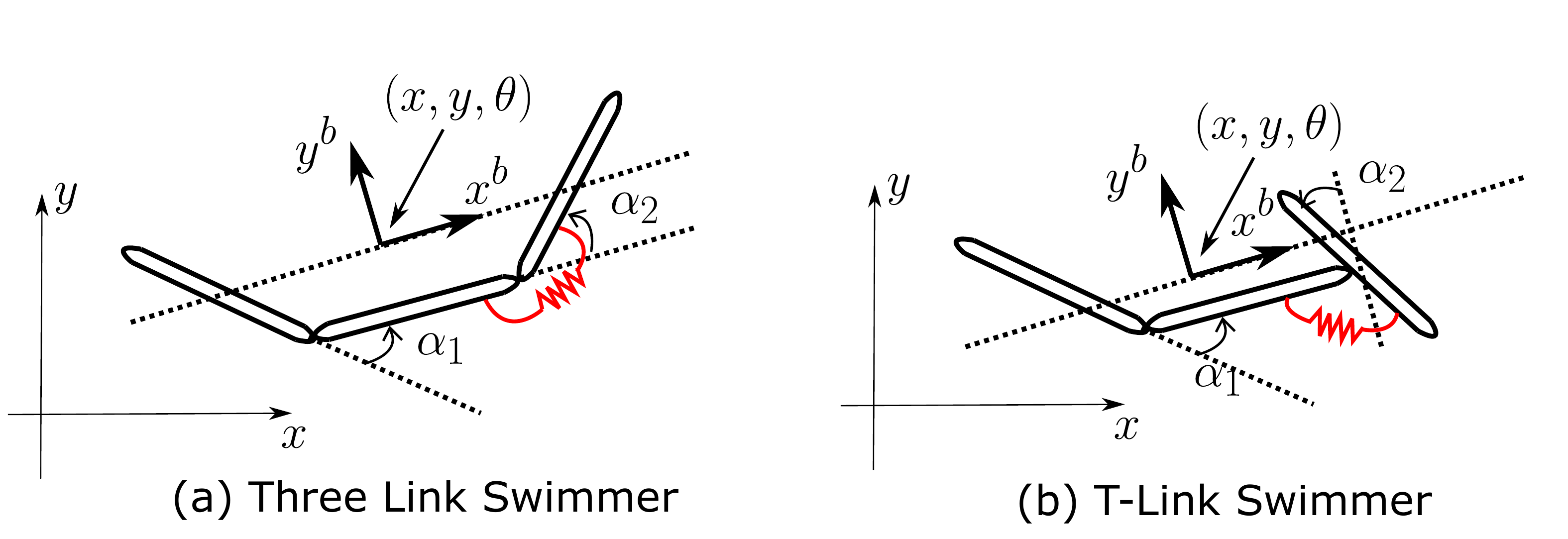}
\caption{The geometry and configuration variables of the Purcell swimmer and T-link swimmer}
\label{fig:3link}
\end{figure}

The Purcell swimmer illustrated in Fig.~\ref{fig:3link}(a) is a classic example three-link system with minimal complexity that can swim in a low Reynolds number fluid (drag-dominated environment), and was introduced in~\cite{Purcell:1977}.

The T-link swimmer illustrated in Fig.~\ref{fig:3link}(b) is a modification of the Purcell swimmer where one of the peripheral links is attached to the center link at its midpoint. The T-link swimmer was introduced in~\cite{krishnamurthy2017schistosoma} as a simplified model to study the swimming dynamics of \textit{Schistosoma mansoni}. \textit{S.mansoni} causes schistosomiasis, a disease comparable to malaria in socio-economic impact~\cite{krishnamurthy2017schistosoma}.

In the passive Purcell swimmer, a torsional spring is attached at the second joint as shown in Fig.~\ref{fig:3link}(a) such that when the joint angle $\alpha_2$ is zero, the spring is at its equilibrium. In the passive T-link swimmer, the torsional spring is attached between the middle link and the peripheral link attached at its midpoint as shown in~\ref{fig:3link}(b) such that when the joint angle $\alpha_2$ is zero, the spring is at its equilibrium.

\subsection{Geometric Locomotion Model}
\label{sec:geo}

When analyzing a locomoting system, it is convenient to separate its configuration space $\bundlespace$ (i.e.\ the space of its generalized coordinates $\bundle$) into a position space $\fiberspace$ and a shape space $\basespace$, such that the position $\fiber\in\fiberspace$ locates the system in the world, and the shape $\base\in\basespace$ gives the relative arrangement of the particles that compose it~\footnote{In the parlance of geometric mechanics, this assigns $\bundlespace$ the structure of a (trivial, principal) \emph{fiber bundle}, with $\fiberspace$ the \emph{fiber space} and $\basespace$ the \emph{base space}.}.

\begin{figure}%
\centering
\includegraphics[width=.45\textwidth]{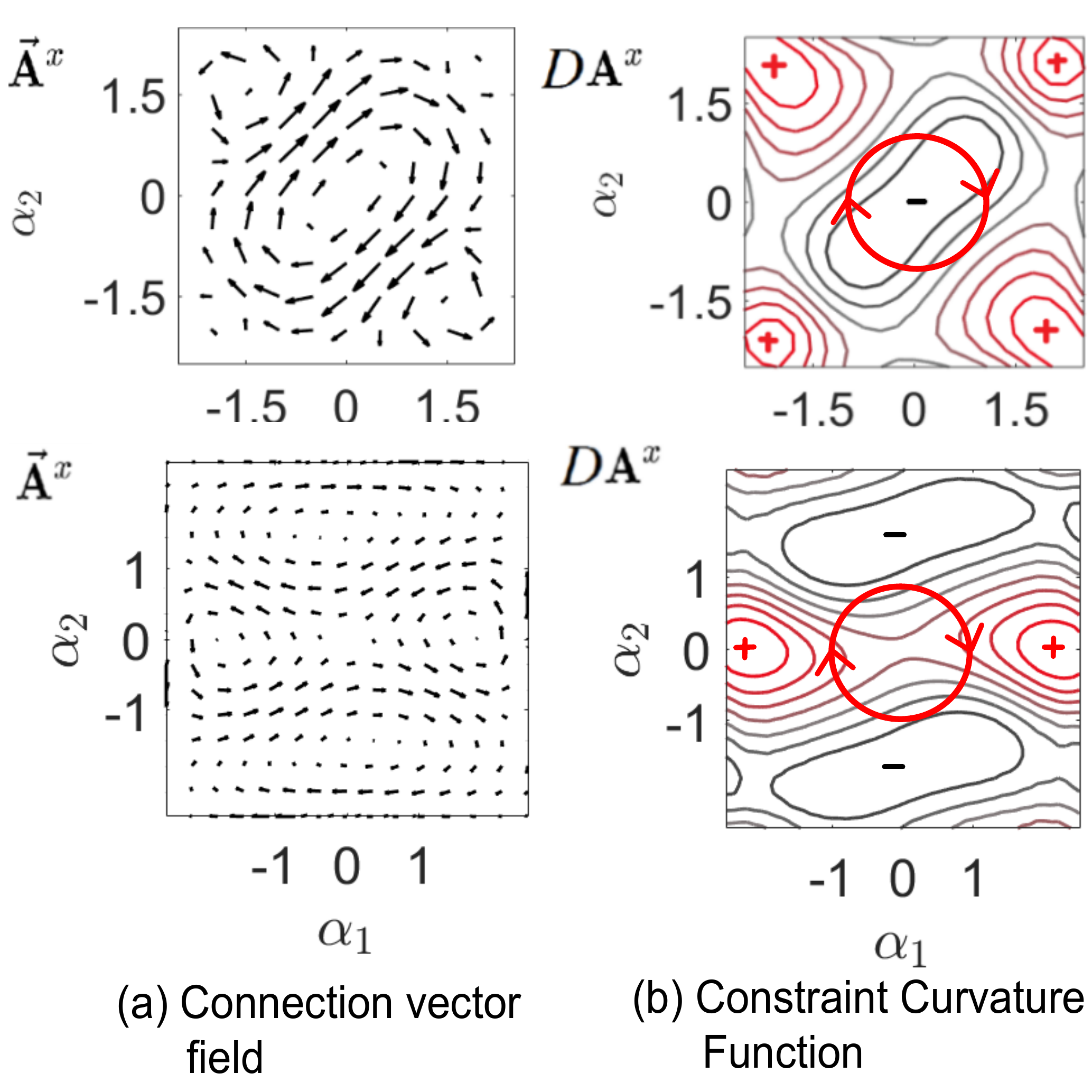}
\caption{The connection vector fields in the x-direction for the Purcell swimmer and T-link swimmer are shown in (a) and the corresponding constraint curvature functions are shown in (b). \cite{Hatton:2013TRO:Swimming}.}
\label{fig:connection3link}
\end{figure}

The locomotion model we employ in this paper was developed for systems in \emph{kinematic} regimes where no gliding can occur i.e, where zero shape velocity results in zero position space velocity~\footnote{This kinematic condition has been demonstrated for a wide variety of physical systems, including those whose behavior is dictated by conservation of momentum~\cite{walsh95, Shammas:2007}, non-holonomic constraints such as passive wheels~\cite{Murray:1993, ostrowski98a, Bloch:03,Shammas:2007}, and fluid interactions at the extremes of low~\cite{Avron:2008,Hatton:2013TRO:Swimming} and high~\cite{Melli:2006,Kanso:2009sw,Hatton:2013TRO:Swimming} Reynolds numbers.}. In this model, there exists a linear relationship at each shape between changes in the system's shape and changes in its position,
\beq
\bodyvel = -\mixedconn(\base) \basedot,
\label{eq:kinrecon}
\eeq
in which $\bodyvel=\fiber^{-1}\fiberdot$ is the body velocity of the system ({i.e.}, $\fiberdot$ expressed in the system's local coordinates), and the \emph{local connection} $\mixedconn$ linearly maps joint velocities to the  body velocity they produce by pushing the system against its environment. Each row of $-\mixedconn$ can be regarded as a body-coordinates local derivative of one position component with respect to the system shape. If we plot the rows of $-\mixedconn$ as arrow fields, as in Fig.~\ref{fig:connection3link}, this means that moving in the direction of the arrows moves the system positively in the corresponding body direction, and moving perpendicular to the arrows results in no motion in that direction~\cite{Hatton:2011IJRR,Hatton:2015EPJ}.

Several efforts  in the geometric mechanics community (including our own), have aimed to use the structure of the systems' \emph{Lie brackets} (a measure of how ``non-canceling" the system dynamics are over cyclic inputs) to understand the structure of the optimal solutions to the system equations of motion~\cite{Kelly:1995,walsh95,ostrowski98a,Melli:2006,Shammas:2007,Avron:2008,Hatton:2013TRO:Swimming,TRO-Cartography}. The core principle in these works is that because the net displacement $\gaitdisp$ over a gait cycle~$\gait$ is the line integral of~\eqref{eq:kinrecon} along $\gait$, the displacement can be approximated~\footnote{This approximation (a generalized form of Stokes' theorem) is a truncation of the Baker-Campbell-Hausdorf series for path-ordered exponentiation on a noncommutative group, and closely related to the Magnus expansion~\cite{Radford:1998,Magnus:1954vl}. For a discussion of the accuracy of this approximation and its derivation, see~\cite{Hatton:2015EPJ}.} by an integral of the curvature $D(-\mixedconn)$ of the local connection (its total Lie bracket, which is a measure of how much the connection changes, over a surface $\gait_{a}$ bounded by the cycle)~\cite{Hatton:2015EPJ}:
\begin{align}
\gaitdisp &= \ointctrclockwise_{\gait} - \fiber\mixedconn(\base) \label{eq:gaitpathintegral}\\ &\approx \exp\iint_{\gait_{a}} \underbrace{-\extd\mixedconn + \sum\limits_{i<j} \big{[}\mixedconn_{i},\mixedconn_{j}\big{]}}_{\text{$D(-\mixedconn)$ (total Lie bracket)}}, \label{eq:lie}
\end{align}
in which $\extd\mixedconn$ is the exterior derivative of the local connection (its generalized row-wise curl),
\begin{equation}
\extd \mixedconn =\Big({\partial \mixedconn_j \over \partial r_i}-{\partial \mixedconn_i \over \partial r_j}\Big) dr_i \wedge dr_j,
\end{equation}
which measures how $\mixedconn$ changes across the shape space and the local Lie bracket term evaluates (on $SE(2)$) as
\beq
\big{[}\mixedconn_{i},\mixedconn_{j}\big{]} = \begin{bmatrix} \mixedconn^{y}_{i}\mixedconn^{\theta}_{j}- \mixedconn^{y}_{j}\mixedconn^{\theta}_{i} \\ \mixedconn^{x}_{j}\mixedconn^{\theta}_{i}- \mixedconn^{x}_{i}\mixedconn^{\theta}_{j} \\ 0 \end{bmatrix}d r_{i} \wedge d r_{j}.
\eeq
which measures how forward and turning motions combine into lateral motions via ``parallel parking".%

In~\cite{Hatton:2010ICRA:BVI,Hatton:2011IJRR,Hatton:2015EPJ}, we identified coordinate choices that make the approximation in~\eqref{eq:lie} accurate for large-amplitude gaits, which allowed us to establish a geometric framework to identify and compare gaits that maximized displacement for fully actuated swimmers in drag-dominated environments.

For systems with just two shape variables, plotting the coefficients of the curvature terms as scalar functions on the shape space reveals the attractors that influence the optimal gait cycles: Gaits that produce net displacement in a given $(x,y,\theta)$ direction encircle strongly sign-definite regions of the corresponding $D(-\mixedconn)$ constraint curvature. For example, as illustrated in Fig.~\ref{fig:connection3link}(b), $x$-translation gaits encircle the center of the shape space for the Purcell system. In this paper as in~\cite{ramasamy2019geometry}, we seek gaits that maximize speed and efficiency of motion in the $x$ direction.

Because the scale of motions expected from these systems are generally much larger than displacements produced by executing one gait cycle, the near-optimal way for these systems to move from one point to another is to orient themselves towards the goal and execute gaits that optimize motion in the body $x$ direction. From hereon, $g_{\phi}$ will refer to the displacement produced in the $x$ direction by the gait $\phi$.

\subsubsection*{System dynamics using RFT}

In this paper, as in~\cite{Hatton:2013TRO:Swimming,TRO-Cartography}, we generate the dynamics for our example systems from a resistive force model, in which each element of the body is subject to normal and tangential drag forces proportional to their velocities in those (local) directions~\footnote{This model is most widely associated with swimmers at low Reynolds numbers (e.g.,~\cite{Tam:2007}),but can also be regarded as an informative general model for systems that experience more lateral drag than longitudinal drag (e.g.~\cite{Hatton:2013PRL}). Our choice of resistive force here also does not preclude the use of more dejointed physical models (e.g.,~\cite{giuliani2018predicting}) to construct the local connection $\mixedconn$.}. The normal drag coefficient is larger than the tangential component (here, by a factor of $2:1$), corresponding to the general principle that it is harder to move a slender object in a fluid or on a surface crosswise than it is to move it along its length~\footnote{Note that a more complete fluid model for low Reynolds number swimming including interbody flow interactions would change numbers in the dynamics but not the overall structure.}.

We combine this linear drag condition with a quasi-static equilibrium condition that the net drag force and moment on the system is zero at all times (treating the system as heavily overdamped, with acceleration forces much smaller than drag forces). Because the drag forces are not isotropic, the quasi-static condition does not prevent the system from moving, and the system can use the angle-of-attack of its body surfaces to generate net motion.

The quasistatic condition imposes a \emph{Pfaffian constraint}~\footnote{A constraint that the allowable velocities are orthogonal to a set of locally-linear constraints, i.e., that they are in the nullspace of a constraint matrix $\omega$.} on the system's generalized velocity, which states that combinations of body and shape velocities that are consistent with the system physics must lie in the null space of a linear map from the system velocities to the forces and moments acting on the system's body frame,
\beq
\begin{bmatrix} F^{b}_{x} \\ \vspace{2pt} F^{b}_{y} \\ F^{b}_{\theta} \end{bmatrix} = \begin{bmatrix} 0 \\ 0 \\ 0 \end{bmatrix} = \omega(\base) \begin{bmatrix} \bodyvel \\ \dot{\base} \end{bmatrix}.\label{eq:forcepfaffian}
\eeq
The matrix $\omega$ that maps the velocities to the net forces on the body frame is a pullback of the drag matrices of the individual links via the system kinematics and depends only on the shape $\base$.

By separating $\omega$ into two sub-blocks, $\omega = [ \omega_{\fiber}^{3\times3}, \omega_{\base}^{3\times n}]$, it is straightforward to rearrange \eqref{eq:forcepfaffian} into
\beq
\bodyvel = -(\omega_{\fiber}^{-1}\omega_{\base})\basedot,\label{eq:invomega}
\eeq
revealing the local connection as $\mixedconn = \omega_{\fiber}^{-1}\omega_{\base}$. The expressions for the dynamics are unwieldy (running to several pages of trigonometric terms in even the simplest cases) so we do not write them out in full here. See~\cite{Hatton:2013TRO:Swimming} for a more detailed treatment of~\eqref{eq:forcepfaffian}--\eqref{eq:invomega} in the case of the three-link swimmer).%

Using the local connection $\mixedconn$ and the system's internal kinematics, we can obtain the Jacobian from shape velocity to the local velocity of each section of the body $J(r,\ell)$, where $\ell$ is the location of the section on the body. We can use this Jacobian to calculate a Riemannian metric $\metric$ over the shape space that encodes the cost of effecting a shape change as
\beq
\metric(\base) = \int_{\text{body}} \Big(\transpose{J}(\base,\ell)\Big)\ C\ \Big(J(\base,\ell) \Big)\ d\ell, \label{eq:metrcdef}
\eeq
where $C$ is the matrix of drag coefficients, which acts as a local metric for the motion of each element of the body. For the systems considered in the paper, we take this matrix as,
\begin{equation}
C=\begin{bmatrix}
2 & 0 & 0 \\
0 & 1 & 0 \\
0 & 0 & 0
\end{bmatrix}
\end{equation}
indicating that for any infinitesimal element of a link, the resistance to lateral motion is twice the resistance to longitudinal motion and that the moment-drag on a link is generated by the translation of the longitudnal and lateral forces acting in the local frame of the infinitesimal element to the body frame of the system~\footnote{This metric corresponds to approximating the viscous drag acting on the swimmer via a local resistive force model. If we used a more accurate fluid model, the drag on a low-Reynolds number swimmer would have the same form, except that $C$ would also depend on $r$ and $\ell$.}.

As discussed in our previous work~\cite{Hatton:2013TRO:Swimming,Hatton:2015EPJ}, the metric $\metric$ encodes a quadratic relationship between the shape velocities and power dissipated into the surroundings, given by
\begin{equation}
P = \basedot^T \metric(r) \basedot \label{eq:costeq},
\end{equation}
as well as the mapping from joint velocities to torques on the joints,
\begin{equation}
\tau=\metric(r)\basedot. \label{eq:torquerel}
\end{equation}

Details of the calculations to generate the local connection $\mixedconn$ and the Riemmanian metric $\metric$ for our example systems are provided in~\cite{Hatton:2013TRO:Swimming} and~\cite{TRO-Cartography}.

In drag-dominated environments, a common measure of the cost of any motion executed by a swimmer is the energy dissipated into the surrounding fluid while executing the motion,
\begin{equation}
E=\int_0^T P(t) dt \label{eq:energyeq}
\end{equation}
Substituting the expression for power dissipated from~\eqref{eq:costeq} provides a measure of the cost in terms of the metric and shape trajectory,
\begin{equation}
E=\int_0^T P(t) dt=\int_0^T \basedot^T \metric(r) \basedot dt.
\end{equation}
This cost depends on the geometry, time period and pacing of a given gait. As discussed in ~\cite{Becker:2003,Avron:2004kx,Hatton:2017TRO:Cartography}, for fully actuated swimmers this cost can be broken down into a combination of a pacing-invariant cost that measures the pathlength $s$ of the trajectory through the shape space (weighted by the shapespace metric $\metric$), and a pacing cost $\sigma$ that measures the deviation from optimal pacing. Finding a gait that minimizes the energy dissipated into the surrounding fluid $E$ is thus equivalent to finding a gait the minimizes the metric-weighted pathlength $s$, and executing said gait at a constant-power pacing to minimize $\sigma$.

The pathlength $s$ of a curve $r(t)$ under a metric $\metric$ is
\begin{equation}
s=\oint ds=\oint \sqrt{dr^T\metric(r)dr}. \label{eq:distmetric}
\end{equation}
Changing the variable of integration from shape $r$ to time $t$, we can relate the pathlength to the square root of power expended:
\begin{equation}
s=\int_0^T\sqrt{\basedot^T \metric(r) \basedot}dt=\int_0^T \sqrt{P(t)} dt \label{eq:energy1}
\end{equation}

Because moving with constant power is the least-costly pacing with which to execute a motion under viscous drag~\cite{Becker:2003}, we can further simplify~\eqref{eq:energy1} as
\begin{equation}
s=\sqrt{P_{\textrm{avg}}}T,
\end{equation}
where $P_{\textrm{avg}}$ is the average power utilized while executing the motion. This pathlength provides a geometric cost for the best-case execution of the kinematics in a gait cycle.

The additional cost for a non-optimal pacing can be represented by squaring the difference between the average and instantaneous rates at which the gait is being followed (measured as $\alnth$ per time), and then integrating over the time during which the gait is being executed,
\beq \label{eq:intstress}
\sigma = \int_{0}^{\tau_{total}} \left(\frac{\alnth_{\text{total}}}{\altlnth_{\text{total}}} - \frac{d}{d\altlnth}(\alnth(\tau))\Big|_{\tau=t} \right)^{2}dt,
\eeq
where $\tau_{total}$ is the time period of the gait, $\alnth_{total}$ is the length of the gait under the metric $\metric$, and $\alnth$ is distance traveled along the gait as a function of time corresponding to the given pacing. If the gait is proceeding at constant power, $\frac{\alnth_{\text{total}}}{\altlnth_{\text{total}}}$ is equal to the rate at which $s$ changes with time, so $\sigma$ measures the extent to which the pacing lags and leads the optimal pacing.

Any pacing other than constant-power will make the trajectory take longer for a given average power (or increase the average power required to complete the motion in a fixed time).

\section{Frequency Domain Analysis}
\label{sec:frequency}

The key difference in the dynamics of a swimmer with a passive joint when compared to a fully actuated swimmer is the coupling of the motion of the actuated and unactuated joint. In this section, we explore this difference further and present a way of accurately approximating the motion of the unactuated joint from the motion of the actuated joint using frequency domain analysis. The method of linearizing the passive dynamics to obtain approximate limit cycles presented in this section is in the same vein as the limit cycle analysis presented in~\cite{Burton:2010}, where a two-link system with static separation between centers of mass and buoyancy was studied.

\subsection{Dynamics of the passive elastic joint}
As discussed in~\cite{Hatton:2013TRO:Swimming,Hatton:2015EPJ}, the mapping between joint velocities and torques on the joint in the fully actuated swimmers is encoded in the metric calculated in~\eqref{eq:metrcdef},
\begin{equation}
\tau=\metric(r)\basedot.
\end{equation}
Because the systems considered in this paper have only one active and one passive joint, this relationship becomes
\begin{equation}
\begin{bmatrix}
\tau_1 \\
\tau_2
\end{bmatrix} = \metric(\alpha_1,\alpha_2) \begin{bmatrix}
\dot{\alpha}_1 \\
\dot{\alpha}_2
\end{bmatrix}.
\end{equation}
In the case of the swimmers with an elastic joint, because the actuation in the second joint is replaced by an elastic element with stiffness $k$, the torque $\tau_2$ is always equal to $-kr_2$, i.e.,
\begin{equation}
\begin{bmatrix}
\phantom{-}\tau_1 \\
-k \alpha_2
\end{bmatrix} = \metric(\alpha_1,\alpha_2) \begin{bmatrix}
\dot{\alpha}_1 \\
\dot{\alpha}_2
\end{bmatrix}.
\end{equation}
The first equation in this system of equations,
\begin{equation}
\tau_1=\metric_{11}\dot{\alpha}_1+\metric_{12}\dot{\alpha}_2,
\end{equation}
thus relates the torque in the actuated joint to the motion of the joints and can be used to calculate the torque required to effect any feasible motion.
The second equation in this system of equations,
\begin{eqnarray}
-k \alpha_2-\metric_{22}\dot{\alpha}_2=
\metric_{21}\dot{\alpha}_1
\end{eqnarray}
or equivalently,
\begin{eqnarray}
-{k \over \metric_{21}} \alpha_2-{\metric_{22} \over \metric_{21}}\dot{\alpha}_2=\dot{\alpha}_1 \label{eq:springmass}
\end{eqnarray}
encodes the dynamics of the passive elastic joint in terms of the active joint, and thus defines the space of feasible motions.

Since the joint $\alpha_1$ is assumed to be the actuated joint, we have full control of $\dot{\alpha}_1$. The value of $\metric$ depends on $\alpha_1$ and $\alpha_2$. Its dependence on $\alpha_1$ and $\alpha_2$ conveys how the shape of the robot affects the effort required to move the joints. If we consider gaits that are relatively small oscillations of shape, we can approximate the value of $\metric$ to be constant throughout the gait. This assumption necessarily introduces errors in our prediction of the motion of the passive joint when the amplitude of input to the active joint is large.

\begin{figure}%
\centering
\includegraphics[width=.37\textwidth]{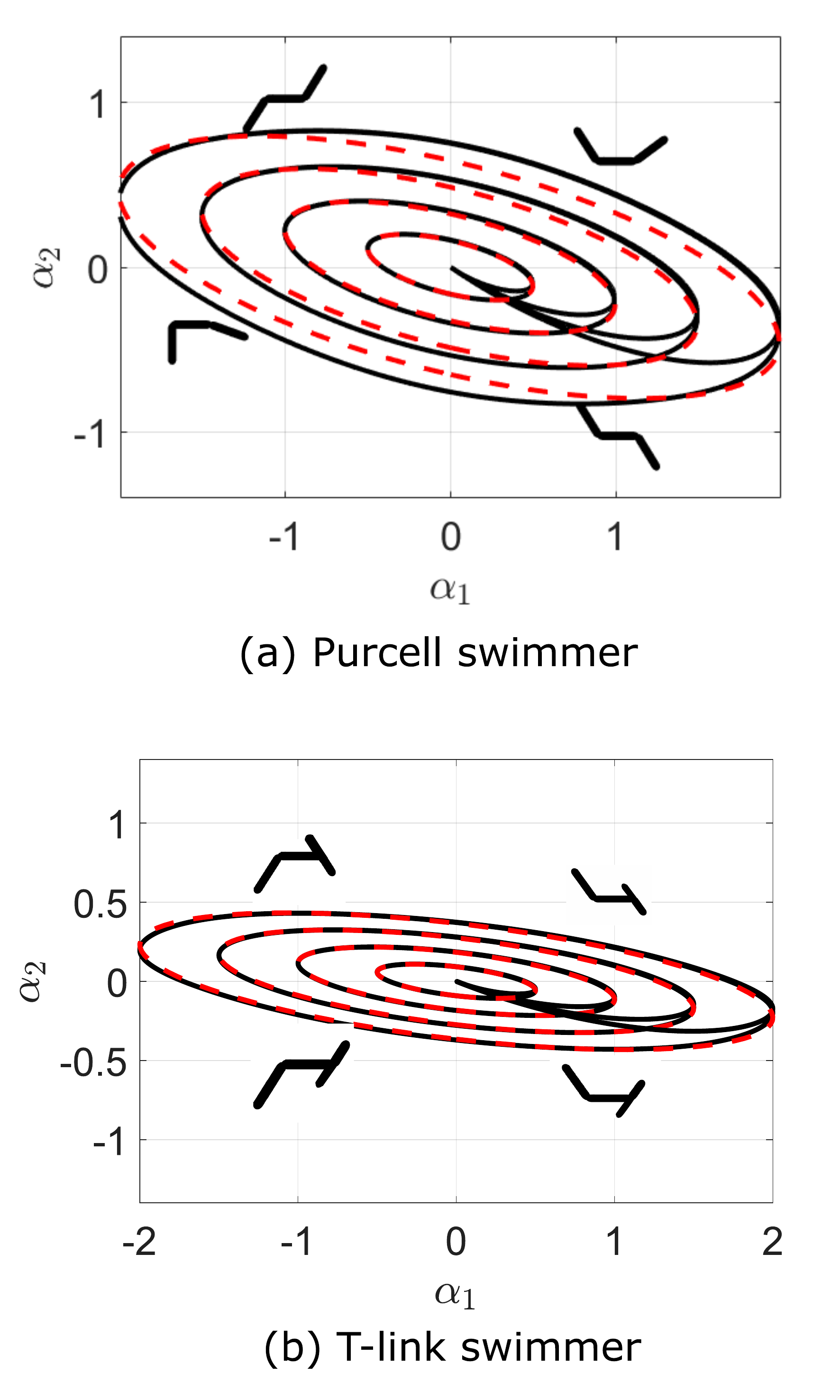}
\caption{Comparison of the exact stroke limit cycles with the shape predicted by frequency domain analysis. Each black solid line represents the motion of the swimmer when the input to the actuated joint is a sinusoidal wave. We can see the motion converges to a limit cycle. The red dashed line represent the shape of the limit cycles predicted by frequency domain analysis presented in \S\ref{sec:frequency}. The cartoons in the background show how the Purcell swimmer looks like at different points of the shape space.}
\label{fig:freqfig}
\end{figure}

For both the Purcell and T-link swimmers with passive elastic joints, assuming the value of $\metric$ to be constant does not introduce significant errors for gaits of amplitude up to 1.5 radians: In Fig.~\ref{fig:freqfig}, we illustrate the distortions caused in the shape of the limit cycles when we assume $\metric$ to be constant throughout the shape space. Each solid black line represents the motion of the full swimmer model when the input to the actuated joint is a sinusoidal wave, and the system starts with both angles at zero. There is a transient term that dominates before the system reaches the limit cycle. The red dashed lines represent the shape of the limit cycles predicted when we assume $\metric$ is constant (we describe the calculation of the limit cycle in the next subsection).

\subsection{Transfer function analysis}
\begin{figure}%
\centering
\includegraphics[width=.25\textwidth]{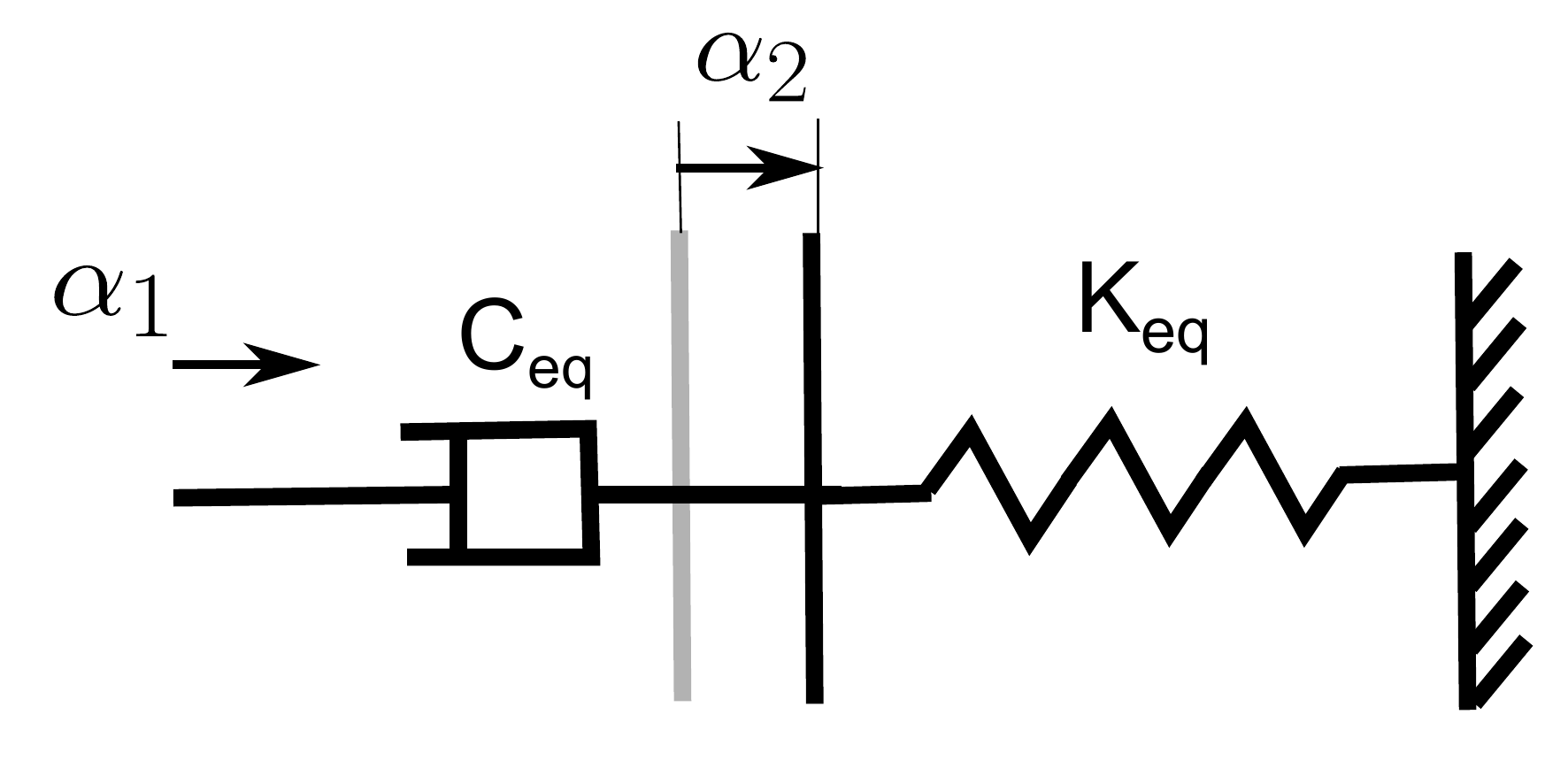}
\caption{Spring damper system whose dynamics are equivalent to the passive dynamics of the Purcell swimmer with a passive elastic joint.}
\label{fig:eqsys}
\end{figure}
In this subsection, our goal is to obtain an analytical approximation of the response of the passive joint to input oscillations of the actively controlled joint. We assume $\metric$ to be constant throughout the gait, which makes~\eqref{eq:springmass}--- the equation that describes the dynamics of the passive elastic joint--- a linear first order differential equation and thus well suited to frequency domain analysis.
\begin{figure*}%
\centering
\includegraphics[width=\textwidth]{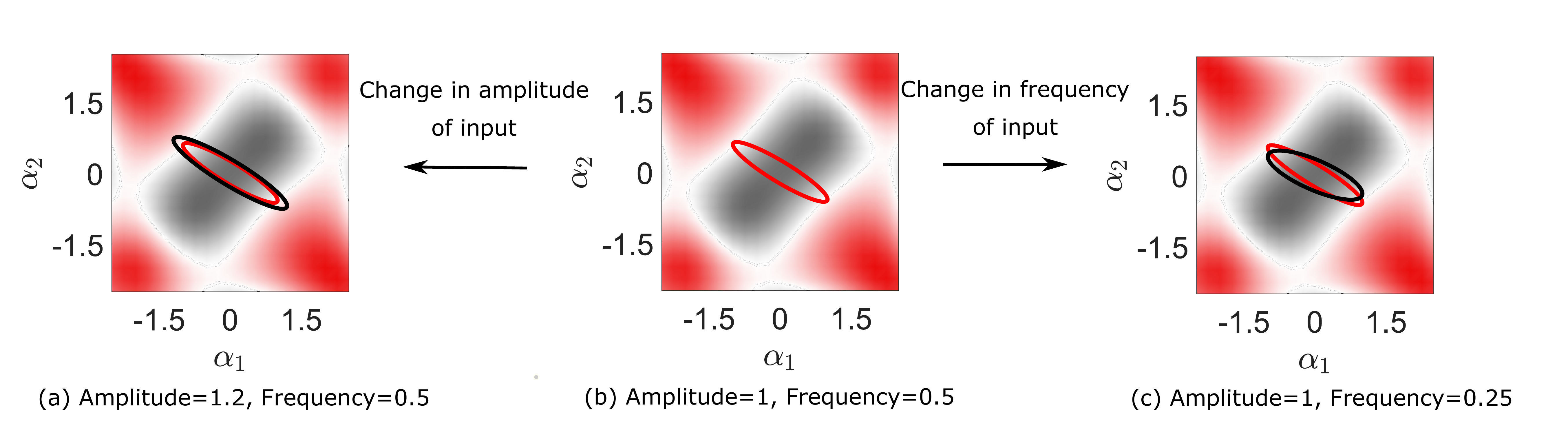}
\caption{Changes in the shape of a gait resulting from changes to amplitude and frequency of input oscillation. (a) The gait resulting from a change in amplitude of the nominal input (black) is a scaled version of the gait corresponding to the nominal input actuation (red). (b) The gait resulting from nominal input actuation. (c) The gait resulting from a change in the frequency of the nominal input (black). A change in the frequency of the nominal input leads to a change in both the amplitude and phase of the response of the passive joint.}
\label{fig:Tandamp}
\end{figure*}

\begin{figure}%
\centering
\includegraphics[width=.45\textwidth]{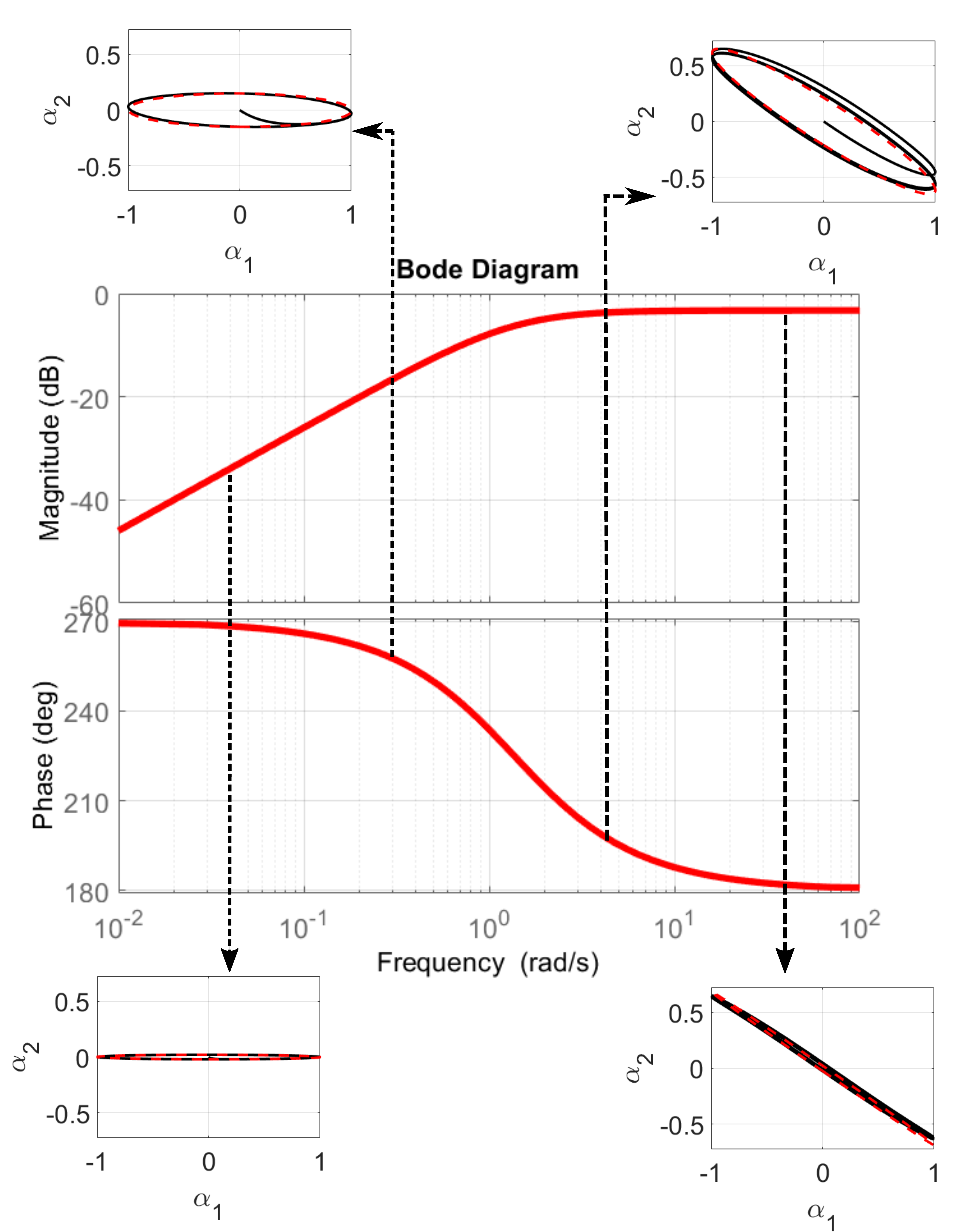}
\caption{Bode plot of the transfer function relating the output of the passive joint to the input of the actuated joint in the Purcell swimmer with a passive elastic tail. The inset figures show how the periodic orbits corresponding to gain and phase at certain frequencies look like in the shape space. We can see that actuation at very low frequencies leads to gaits that enclose very little surface area due to the amplitude of the passive joint being low, while at very high frequencies the surface area enclosed by the gait is low due to the almost 180 degree phase shift between the oscillations of the actuated and passive joints.}
\label{fig:bode}
\end{figure}

We use the Laplace transform on~\eqref{eq:springmass} to obtain the dynamics of the passive elastic joint in the frequency domain as
\begin{eqnarray}
K_{eq}\mathcal{L}(\alpha_2(t))+C_{eq}s\mathcal{L}(\alpha_2(t))=s\mathcal{L}(\alpha_1(t)) \label{eq:eqsys1}
\end{eqnarray}
where \( K_{eq}= \displaystyle{-k \over \phantom{-}\metric_{21}}\) and \( C_{eq}= \displaystyle{-\metric_{22} \over \phantom{-}\metric_{21}}\). These substitutions reveal that the dynamics of the passive joint are equivalent to those of a massless particle attached to a fixed base through a spring and being driven through a damper by a position trajectory $\alpha_1$ as shown in Fig.~\ref{fig:eqsys}. Note that the damping coefficient is completely dependent on the physics of the interaction between the swimmer and the fluid.

We can rewrite~\eqref{eq:eqsys1} as a transfer function relation between the active joint $\alpha_1$ and passive elastic joint $\alpha_2$
\begin{eqnarray}
\mathcal{L}(\alpha_2(t))=H(s)\mathcal{L}(\alpha_1(t)) \label{eq:transfer function}
\end{eqnarray}
where \(H(s)=\displaystyle {s \over (C_{eq}s +K_{eq})}\) is the transfer function that encodes the response of the passive elastic joint to osciallations of the active joint.

Equation~\eqref{eq:transfer function} tells us how inputs to the controlled joint, $\alpha_1$ are mapped to the response of the passive joint in the frequency domain. In order to find the response of the passive joint to a sinusoidal oscillation of the actuated joint, we let $\alpha_1(t)=\sin(wt)$, then using~\eqref{eq:transfer function}, we obtain
\begin{eqnarray}
\mathcal{L}(\alpha_2(t))&=&\overbrace{{s \over (C_{eq}s +K_{eq})}}^{H(s)}\overbrace{{\omega \over (s^2+ \omega^2)}}^{\mathcal{L}(\alpha_1(t))} \label{eq:transfidk}\\
&=&\underbrace{{A_1 \over (C_{eq}s +K_{eq})}}_{\text{Transient term}}\quad+\quad \underbrace{{A_2s +A_3 \over (s^2+ \omega^2)}}_{\mathclap{\substack{\text{Phase shifted sine wave}}}} \label{eq:transf1}
\end{eqnarray}
where $A_1$, $A_2$ and $A_3$ can be obtained by equating the two expressions for $\mathcal{L}(\alpha_2(t))$ in~\eqref{eq:transfidk} and~\eqref{eq:transf1} as
\begin{equation}
s\omega=A_1(s^2+\omega^2)+(A_2s +A_3)(C_{eq}s +K_{eq}),
\end{equation}
and equating the coefficients of powers of $s$ on each side to extract a system of three equations,
\begin{eqnarray}
\begin{bmatrix}
1 & C_{eq} & 0 \\
0 & K_{eq} & C_{eq} \\
\omega^2 & 0 & K_{eq}
\end{bmatrix}\begin{bmatrix}
A_1 \\
A_2 \\
A_3
\end{bmatrix}=\begin{bmatrix}
0 \\
\omega \\
0
\end{bmatrix} \label{eq:transf2},
\end{eqnarray}
which can be easily solved to obtain $A_1$, $A_2$ and $A_3$ as
\begin{equation}
\begin{bmatrix}
A_1 \\
A_2 \\
A_3
\end{bmatrix}=\begin{bmatrix}
1 & C_{eq} & 0 \\
0 & K_{eq} & C_{eq} \\
\omega^2 & 0 & K_{eq}
\end{bmatrix}^{-1}\begin{bmatrix}
0 \\
\omega \\
0
\end{bmatrix}.
\end{equation}

If our actuated joint follows a trajectory given by $\alpha_1(t)=B_1\sin(\omega t)$, after reaching the steady state periodic orbit, the unactuated joint, therefore, follows the trajectory
\begin{equation}
\alpha_2(t)=B_1A_2\cos(\omega t)+B_1A_3\sin(\omega t).
\end{equation}
If our actuated joint follows a trajectory given by $\alpha_1(t)=B_1\sin(\omega t)+B_2\cos(\omega t)$, after reaching the steady state periodic orbit, the unactuated joint follows the trajectory
\begin{eqnarray}
\alpha_2(t)&=&B_1(A_2\cos(\omega t)+A_3\sin(\omega t)) \nonumber \\
&&+B_2(-A_2\sin(\omega t)+A_3\cos(\omega t)) \label{eq:transferfcneqn}.
\end{eqnarray}

Note that the value of $A_2$ and $A_3$ depend on the value of $\omega$, the frequency of the sinusoidal input to the actuated joint. The image of the gait in the shape space is thus coupled to the pacing of the input to the actuated joint.

This coupling between the image of the gait in the shape space and the input to the actuated joint is illustrated in Fig.~\ref{fig:Tandamp}. Fig.~\ref{fig:Tandamp}(b) shows the shape of the gait resulting from a sinusoidal oscillation of the actuated joint of amplitude 1 and frequency 0.5. Fig.~\ref{fig:Tandamp}(a) shows the effect of increasing the amplitude of oscillation on the shape of the gait. As we have linearized the dynamics of the passive joint, a change in amplitude without a change in frequency produces a scaled version of the original gait. Fig.~\ref{fig:Tandamp}(c) shows the effect of decreasing the frequency of oscillation on the shape of the gait, with a slower oscillation leading to a weaker response (in terms of magnitude) from the passive joint.

The Bode plot of the transfer function $H$ in~\eqref{eq:transfer function}, presented in Fig.~\ref{fig:bode}, illustrates the specific nature of the relationship. We can see that if the frequency of input to the controlled joint is very high, the response of the passive elastic joint is phase shifted relative to the actuated joint by almost $180$ degrees, resulting in a gait with a small area. At very low frequencies, the magnitude of the response of the passive elastic joint is small and results in a gait with a small area. At mid-range frequencies, however, the response of the passive elastic joint has a larger amplitude and a phase shift that results in gaits with larger areas.

\section{Finding Optimal Gaits}
\label{sec:optimalgaits}
Optimal gait design has a long history of research in the physics, mathematics, and engineering communities, as part of the broader field of optimal control~\cite{Bryson:1979,Ostrowski:2000vl}. Notable contributions to finding optimal gaits for swimmers in a drag-dominated environment include those of Purcell, who introduced the three-link swimmer as a minimal template for understanding locomotion, a series of works~\cite{Tam:2007,DeSimone:2012aa,Giraldi:2013aa,Bettiol:2015,Shapere:1989} aimed at numerically optimizing the stroke pattern, and the observation in~\cite{Becker:2003} that the optimal pacing for the gait keeps the power dissipation constant over the cycle. %

In drag-dominated environments, a common measure of the cost of any motion executed by a swimmer is the power dissipated into the surrounding fluid while executing the motion. A natural choice of definition for efficiency for these systems is thus
\begin{eqnarray}
\eta_1={g_{\phi} \over E}, \label{eq:eta1}
\end{eqnarray}
where $g_{\phi}$ is the displacement produced in the $x$ direction over a gait cycle $\phi$ and $E$ as defined in~\eqref{eq:energyeq} is the total energy dissipated into the surrounding while executing the gait $\phi$.

Note that this definition of efficiency is the inverse of the mechanical cost of transport used, e.g., in~\cite{Avron:2004kx}. The mechanical cost of transport is a widely used efficiency metric in the robotics community, especially while studying legged locomotion. As explained in Appendix~\ref{appendix:lighthill}, gaits which optimize our criterion also optimize Lighthill's efficiency~\cite{Lighthill:1976}, which compares the power dissipated while executing the gait to the power dissipated in rigidly translating the swimmer through the fluid; this measure has an advantage over Lighthill in that it allows for effective comparison between systems with different morphologies and does not require designating a reference shape for rigidly dragging the swimmer.

As noted in~\S\ref{sec:background} C, under optimal pacing (constant-power pacing), the pathlength $s$ defined in~\eqref{eq:distmetric} and the total power dissipated $E$ are equivalent cost criteria. As detailed in~\cite{ramasamy2019geometry}, for fully actuated swimmers this insight lets us restrict our optimization to constant power trajectories and utilize a geometric definition of efficiency,
\begin{equation}
\eta_2={g_{\phi} \over s} \label{eq:eta2}
\end{equation}

Another important observation to be made here is that, as noted in~\cite{ramasamy2016soap,ramasamy2017geometric}, for the fully actuated swimmers, the gaits that optimized for forward velocity and efficiency were the same within each system. There are two reasons for this:
\begin{packed_enum}
\ie At any given power level $P$, maximum forward speed is attained by executing the efficiency-maximizing gait at a pacing that has an even power usage instead of surging and sagging.
\ie The shape of the efficiency-maximizing gait is invariant across power levels because the displacement resulting from, and the cost incurred by, executing one cycle of any gait at optimal pacing does not depend on the power level $P$.
\end{packed_enum}
The maximum power available, $P$, dictates the time period $T$ of the  speed-maximizing gait. The attainable maximum speed increases linearly with the amount of power available. For efficiency-maximizing gait, the efficiency is invariant across our choice of time period, $T$, because the displacement resulting from and the cost incurred by executing one cycle of any gait at optimal pacing, does not depend on the power level $P$. However, the maximum power available $P$ sets a lower bound on the possible range for $T$.

Moving from an active to a passive swimmer also affects the gradient calculation process that forms the backbone of the framework presented in~\cite{ramasamy2019geometry}. In a passive swimmer, the movement of the passive joint is coupled to the movement of the active joint, and therefore we cannot parametrize the motion of the passive joint independently of the motion of the active joint. We discuss the implications of a passive joint on gait parametrization and gradient calculation in~\S\ref{sec:optimalgaits} B.

\subsection{Efficiency for swimmers with passive elastic joints}
In fully actuated swimmers, the transformation of efficiency from the inverse of cost of transport as defined in~\eqref{eq:eta1} to a more geometric definition in~\eqref{eq:eta2} was possible because we knew that the optimal pacing keeps the rate of power dissipation constant~\cite{Becker:2003}. In the case of swimmers with passive joints, the response of the passive joint is dictated by the dynamics of the active joint, as illustrated by the Bode plots shown in Fig.~\ref{fig:bode}. As a result, there is a unique pacing associated with every gait the passive swimmer can execute. Changing the pacing of the actuated joint changes the response of the passive joint as shown in \S\ref{sec:frequency} and hence the shape of the gait. Thus to find the most efficient gait for the Purcell swimmer with a passive elastic joint, we have to directly use the definition of efficiency in~\eqref{eq:eta1}.

While we could choose a constant power pacing for all gaits for fully actuated swimmers, in the case of the Purcell and T-link swimmers with a passive elastic joint, every gait that respects the passive dynamics of the elastic joint comes with an inherent pacing. Thus for a given spring stiffness of the passive joint, the gait that maximizes forward velocity comes with a power requirement associated with it. Even if we are capable of giving the system more power, there is no way for the system to utilize that power to go faster. Hence for the swimming systems considered in this paper, there are two meaningful measures for comparing different gaits that lead to different definitions of gait optimality: Gaits can be compared by
\begin{packed_enum}
\ie Comparing the average speeds they produce ($\eta= {g_{\phi} \over T}$)
\ie Comparing their energetic efficiency ($\eta_1$ in~\eqref{eq:eta1})
\end{packed_enum}

\subsection{Gait parametrization for passive swimmers}
\label{sec:fourierpara}
We use a truncated Fourier series to parametrize the gaits. This choice of parametrization lets us accurately approximate a large family of smooth periodic gaits. The framework introduced in~\cite{ramasamy2019geometry} uses a gradient descent algorithm to identify gaits that maximize efficiency as defined in~\eqref{eq:eta2}. During the gradient calculation process outlined in appendix~\ref{app:soap}, it is useful to think of the gait as being parametrized by a series of waypoints. We can generate these waypoints from the Fourier parametrization. We use the gradients calculated at each of these waypoints to calculate gradients with respect to the Fourier series parametrization.

In the case of swimmers with a passive joint, we let the actuated joint trajectory $\alpha_1(t)$ be given by a fourth order Fourier series,

\begin{eqnarray}
\alpha_1(t)=a_0+\sum_{i=1}^{4} a_i \cos\Big({2\pi i \over T}t\Big)+ b_i \sin\Big({2\pi i \over T}t\Big) \label{eq:alpha1}
\end{eqnarray}
Using~\eqref{eq:transfer function}, i.e. the transfer function relating the movement of the active and passive joints, we obtain the response of passive joint to $\alpha_1(t)$ as
\begin{eqnarray}
\alpha_2(t)=\mathcal{L}^{-1}(H(s)\mathcal{L}(\alpha_1(t)). \label{eq:alpha2} \label{eq:transf3}
\end{eqnarray}
Using explicit evaluation of the transfer function from~\eqref{eq:transf1} and~\eqref{eq:transf2}, we can write the steady state response of the passive joint as
\begin{eqnarray}
\alpha_2(t)=\sum_{i=1}^{4} c_i \cos\Big({2\pi i \over T}t\Big)+ d_i \sin\Big({2\pi i \over T}t\Big), \label{eq:transf4}
\end{eqnarray}
where $c_i$ and $d_i$ are functions of $a_i$ and $b_i$ and $T$.

Using this low-order Fourier series parameterization of the gait, we can generate the direct transcription waypoints, calculate the gradient of the objective function at each waypoint (details of this gradient calculation process for speed-maximizing and efficiency-maximizing gaits are presented in~\S\ref{sec:Optimizer} and~\S\ref{sec:energyopt} respectively), then project these gradients onto the Fourier basis, to obtain gradients with respect to the Fourier series parameters.

If the system were fully actuated, we could move the Fourier series parameters along these calculated gradient directions to obtain the optimal gait. In the case of passive swimmers, the Fourier coefficients of the unactuated shape direction $(c_i,d_i)$ are functions of the Fourier shape coefficients of actuated shape direction $(a_i,b_i)$. Therefore, to find the correct gradient directions for the Fourier coefficients of the actuated shape, we have to account for the change in the unactuated shape direction that a change in the actuated shape direction would produce.

For an objective function $f$ that depends on $a_i,b_i,c_i$ and $d_i$, we can calculate the total derivatives of $f$ with respect to $a_i$ and $b_i$ as
\begin{eqnarray}
{d f\over d a_i}={\partial f\over \partial a_i}+{\partial c_i \over \partial a_i}{\partial f\over \partial c_i}+{\partial d_i \over \partial a_i}{\partial f \over \partial d_i} \label{eq:fouriergrad1}\\
{d f\over d b_i}={\partial f \over \partial b_i}+{\partial c_i \over \partial b_i}{\partial f\over \partial c_i}+{\partial d_i \over \partial b_i}{\partial f\over \partial d_i}\label{eq:fouriergrad2},
\end{eqnarray}
where ${\partial c_i \over \partial a_i},{\partial c_i \over \partial b_i},{\partial d_i \over \partial a_i}$ and ${\partial d_i \over \partial b_i}$ are directly taken from the transfer function coefficients~\eqref{eq:transferfcneqn}
\begin{eqnarray}
{\partial c_i \over \partial a_i}&=& A_2 \\
{\partial c_i \over \partial b_i}&=& A_3 \\
{\partial d_i \over \partial a_i}&=& -A_2 \\
{\partial d_i \over \partial b_i}&=& A_3 .
\end{eqnarray}

We use these total derivatives to calculate the correct gradient directions for the Fourier coefficients of the actuated shape variables, which account for the fact that in passive swimmers, a change in the shape of input to the actuated shape variable affects the response of the passive elastic joint.

\section{Speed-maximizing Gaits}
\label{sec:Optimizer}

\begin{figure}%
\centering
\includegraphics[width=.49\textwidth]{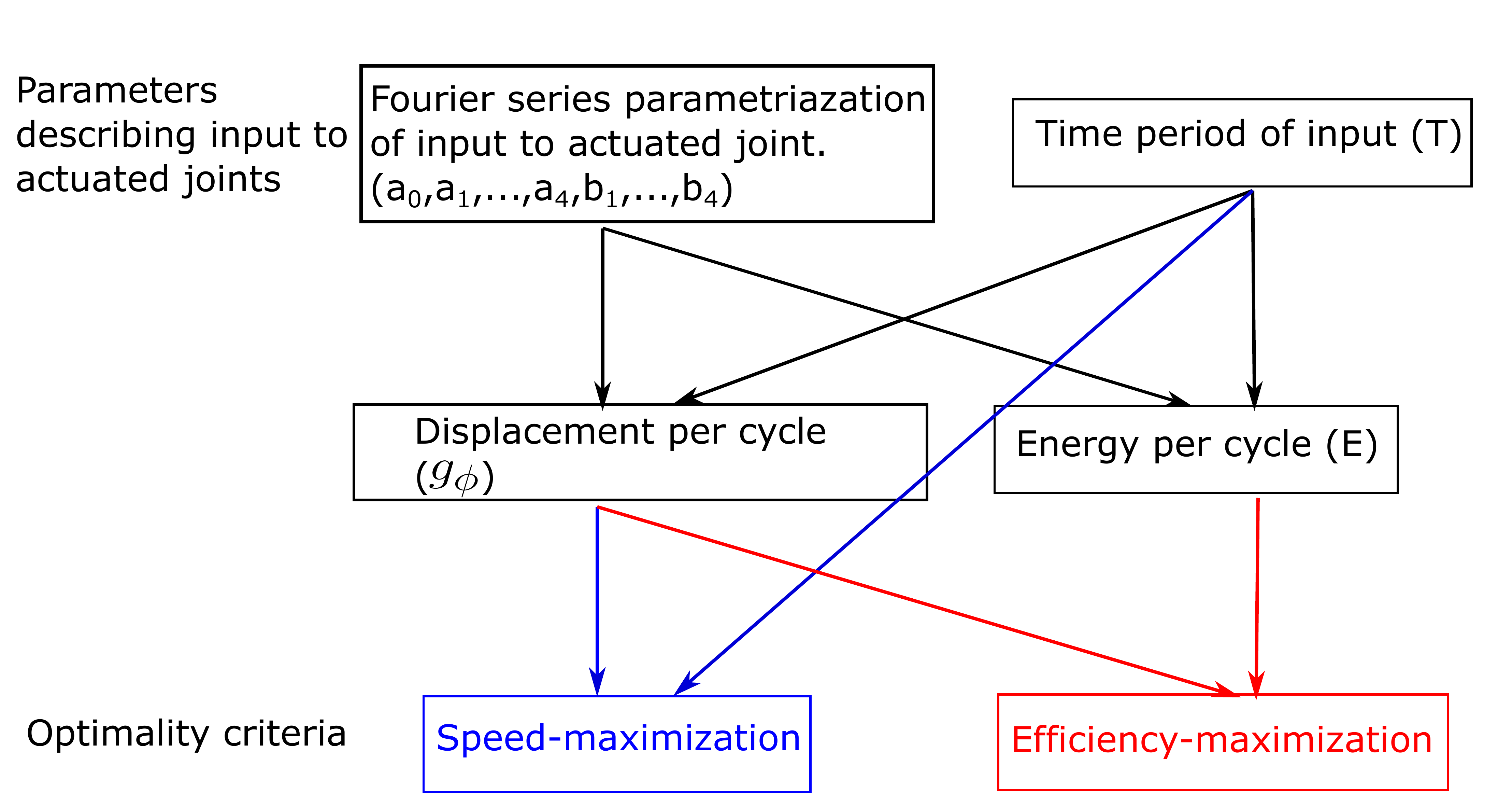}
\caption{A flowchart describing how the parameters describing our input to the actuated joint affect the two optimality criteria in the case of drag-dominated swimmers with a passive joint.}
\label{fig:flow}
\end{figure}
A major goal of this paper is to geometrically identify gaits for the Purcell and T-link swimmers with an elastic joint that will give it the maximum forward velocity. Therefore the objective function we set out to maximize is ${g_{\phi} \over T}$, where $g_{\phi}$ is the displacement in the body $x$ direction produced on executing the gait $\phi$, and $T$ is the time period required to execute the gait.

As explained in~\S\ref{sec:optimalgaits}.B, we parametrize the actuated shape variable using a low-order Fourier series and obtain the Fourier series parametrization of the resulting motion of the passive joint using the dynamics of the passive elastic joint presented in~\S\ref{sec:frequency}. With these Fourier-series parameters, $p_f$, we can obtain a sequence of waypoints $p_i$ (in our implementation we use $100$ waypoints), equally spaced in time, that describe the location of the discretization points in the shape space. As illustrated in this section, we can then calculate the gradient of speed with respect to each of these waypoints, i.e., calculate the effect moving the waypoints would have on the forward speed attained by the swimmer on executing the gait. We then project these gradients onto the Fourier series basis as explained in~\S\ref{sec:optimalgaits}.B to obtain the speed-maximizing gaits using gradient descent.

We start from the basic variational principle that functions reach their extrema when their derivatives go to zero. Given a gait parametrization defined by waypoint parameters $\varparam$, executed in time $T$, the maximum-velocity cycle must satisfy the condition that the gradient of forward velocity with respect to the parameters $\varparam$ and $T$ is zero, i.e.
\begin{eqnarray}
{\partial \over \partial \varparam}\Big({\gaitdisp \over T}\Big)&=&{1 \over T }{\partial \gaitdisp \over \partial \varparam}-{\gaitdisp \over T^2}{\partial T \over \partial \varparam} \label{eq:speedopfirst}\\
&=&{1 \over T }{\partial \gaitdisp \over \partial \varparam} =0,
\end{eqnarray}
and
\begin{eqnarray}
{\partial \over \partial T}\Big({\gaitdisp \over T}\Big)={1 \over T }{\partial\gaitdisp \over \partial T}-{g_{\phi} \over T^2}=0,
\end{eqnarray}
where ${\partial T \over \partial \varparam}$ term in~\eqref{eq:speedopfirst} can be taken as zero since $T$ and $p$ are two independent variables describing the gait, i.e. the time taken to complete a gait $T$ does not depend on the shape and pacing of the input to the actuated joint described by $\varparam$.

Once we have the gradient of speed with respect to a direct transcription parametrization $p$ obtained from a Fourier series parametrization $p_f$, we follow the process outlined in~\S\ref{sec:fourierpara}, specifically~\eqref{eq:fouriergrad1} and~\eqref{eq:fouriergrad2}, to obtain the gradient of speed with respect to the Fourier series parametrization, ${\partial \over \partial \varparam_f}\Big({\gaitdisp \over T}\Big)$. A graphical depiction of how elements of the Fourier series parametrization affect the gait optimization process is shown in Fig.~\ref{fig:flow}.

For suitable seed values $p_{f0}$ and $T_0$, the maximum-velocity gaits can thus be obtained by finding the equilibrium of the dynamical system
\begin{eqnarray}
\dot{p}_f&=&{\partial \over \partial \varparam_f}\Big({\gaitdisp \over T}\Big) \label{eq:fullequation1} \\
\dot{T}&=&{\partial \over \partial T}\Big({\gaitdisp \over T}\Big)={1 \over T }{\partial \gaitdisp \over \partial T}-{g_{\phi} \over T^2}. \label{eq:fullequation2}
\end{eqnarray}

In the geometric framework we introduced in~\cite{ramasamy2016soap}, we showed that the process of finding efficient gaits for the fully actuated Purcell swimmer is akin to the dynamics of a soap-bubble in which internal pressure and surface tension combine to determine the shape, size and surface concentration of the soap bubble. This is not the case in swimmers with a passive elastic joint. Instead of two independent processes, the optimization process for finding the fastest gait is more unified where~\eqref{eq:fullequation} is the equation that helps obtain the shape of the optimal input to the actuated joint and~\eqref{eq:fullequation2} is the equation that helps obtain the optimal pacing of the input.%

\subsection{Shape gradient of the optimal input to the actuated joint.}
\label{subsec:Shapespeed}

As discussed in~\cite{ramasamy2019geometry}, the gradient that affects the shape of the input to the actuated joint, ${d \gaitdisp \over d \varparam}$, pushes the gait towards maximum displacement cycles. From~\eqref{eq:lie}, and the fact that variations in $\varparam$ affect $\gait$ but not the underlying $D\mixedconn$ structure, we can reduce this term to
\beq\label{eq:surfaceint}
{\partial \gaitdisp \over \partial \varparam} \approx \nabla_{\varparam}\iint_{\gait_{a}}{(-D\mixedconn)}.
\eeq
A powerful geometric principle (the general form of the \emph{Leibniz integral rule}~\cite{Flanders:1973aa}) tells us that
\emph{
the gradient of an integral with respect to variations of its boundary is equal to the gradient of the boundary with respect to these variations, multiplied by the integrand evaluated along the boundary,
} and allows us to rewrite \eqref{eq:surfaceint} in terms of the constraint curvature value and how changes in parameter values move the gait's trajectory through the shape space.

Formally, the multiplication of the gradient of the boundary and the integrand evaluated along the boundary is the \emph{interior product}~\footnote{Not the inner product, see~\cite{Flanders:1973aa} for more details.} of the boundary gradient with the integrand,
\beq \label{eq:interiorproduct}
\nabla_{\varparam}\iint_{\gait_{a}}{D(\solneg\mixedconn)} = \ointctrclockwise_{\gait} (\nabla_{\varparam}\gait) \intprod D(\solneg\mixedconn).
\eeq

In systems with just two shape variables, the interior product reduces to a simple multiplication between the outward component of $\nabla_{\varparam}\gait$ and the scalar magnitude of the Lie bracket,
\beq \label{eq:boundarygradientsimple}
\nabla_{\varparam}\iint_{\gait_{a}}{(-D\mixedconn)} = \oint_{\gait} (\nabla_{\varparam_\perp}\phi) (-D\mixedconn).
\eeq

This gradient calculation is illustrated in Fig.~\ref{fig:derivation}. The gradient of the enclosed area with respect to variations in the position of $\varparam_{i}$, i.e. $\nabla_{\varparam_{i}}\gait_{a}$ in the~$\localbasis_{\parallel}$ and~$\localbasis_{\perp}$ directions, is the change in triangle's area as $\varparam_{i}$ moves. Because the triangle's area is always one half base times height (regardless of its pitch or the ratio of its sidelengths), this gradient evaluates to
\beq
\nabla_{\varparam_{i}}\gait_{a} =  \begin{bmatrix} \localbasis_{\parallel} & \localbasis_{\perp} \end{bmatrix}\begin{bmatrix} 0 \\ \ell/2 \end{bmatrix}.
\eeq
Note that this term matches the right-hand side of~\eqref{eq:boundarygradientsimple}, with only normal motions of the boundary affecting the enclosed area.

\subsection{Frequency gradient of the optimal input to the actuated joint}
\label{subsec:freqspeed}
In the case of the fully actuated Purcell swimmer, the shape of the gait, and therefore the displacement produced by executing the gait, are independent of the time taken to execute the gait. This is not true in the case of the Purcell swimmer with a passive elastic joint.

In this subsection, we examine the gradient that guides the optimizer towards the optimal frequency of input to the actuated joint. When the time period required to execute the gait is changed, the shape of the gait changes due to the coupling between the frequency of input to the actuated joint and the response of the passive joint as described in \S\ref{sec:frequency}. Changing the time period $T$ thus changes not only the frequency of the gait cycle but also the displacement produced per cycle.

We use the chain rule to calculate this gradient,

\begin{eqnarray}
{\partial \over \partial T}\Big({\gaitdisp \over T}\Big)&=&{1 \over T }{\partial \gaitdisp \over \partial T}-{g_{\phi} \over T^2} \\
&=&{1 \over T}\Bigg({\partial \gaitdisp \over \partial \alpha_1}{\partial \alpha_1 \over \partial T}+{\partial \gaitdisp \over \partial \alpha_2}{\partial \alpha_2 \over \partial T}\Bigg)-{g_{\phi} \over T^2}
\end{eqnarray}
Because $\alpha_1$ is the actuated shape variable and the shape of the input actuation is independent of the frequency of actuation, ${\partial \alpha_1 \over \partial T}$ reduces to zero. Therefore, the gradient of speed with respect to $T$ reduces to
\begin{eqnarray}
{\partial \over \partial T}\Big({\gaitdisp \over T}\Big)&=& {1 \over T}\Bigg({\partial \gaitdisp \over \partial \alpha_2}{\partial \alpha_2 \over \partial T}\Bigg)-{g_{\phi} \over T^2}. \label{eq:speedop2}
\end{eqnarray}
The first term of the right hand side of~\eqref{eq:speedop2} captures the contribution to the velocity of the gait caused by the change in the shape of the gait resulting from a change in $T$. The second term accounts for the fact that, even without a change in the shape of the gait, an increase in the time required to execute the gait would result in a decrease in the velocity of the gait.

\begin{figure*}%
\centering
\includegraphics[width=\textwidth]{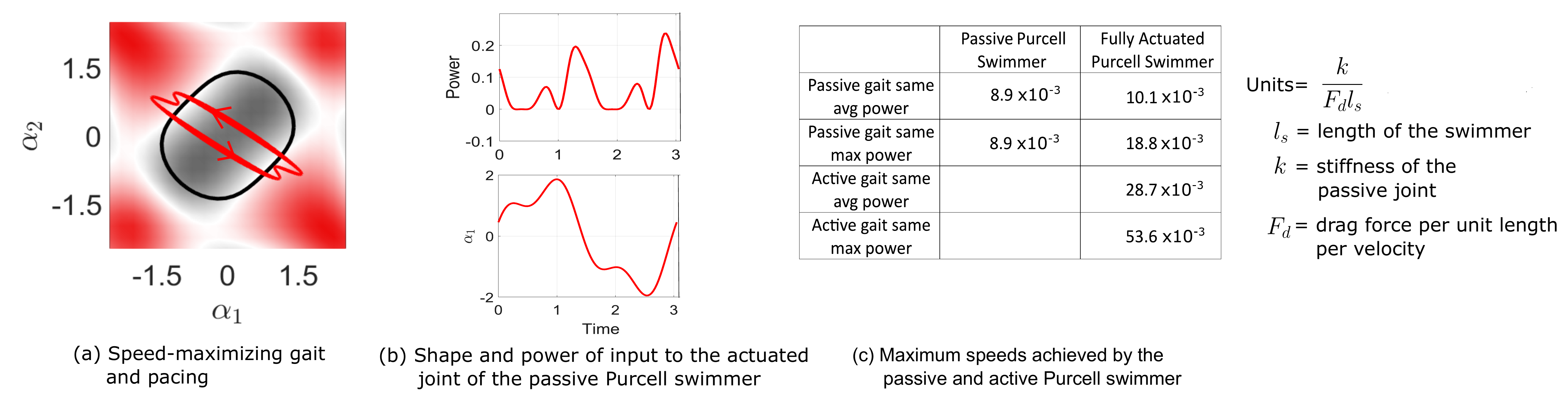}
\caption{Gaits that maximize speed along x-direction for the Purcell swimmers. (a) The optimal gait for the passive Purcell swimmer (red) and the optimal gait for the fully actuated Purcell swimmer (black). The thickness of the line shows the magnitude of power required at different points of the gait. (b) The power required to execute the optimal gait for the passive Purcell swimmer and input to the actuated joint over one gait cycle. (c) A comparison table of the speeds achieved by the passive and fully actuated Purcell swimmers at different power levels.}
\label{fig:maxspeed3}
\end{figure*}

\begin{figure*}%
\centering
\includegraphics[width=\textwidth]{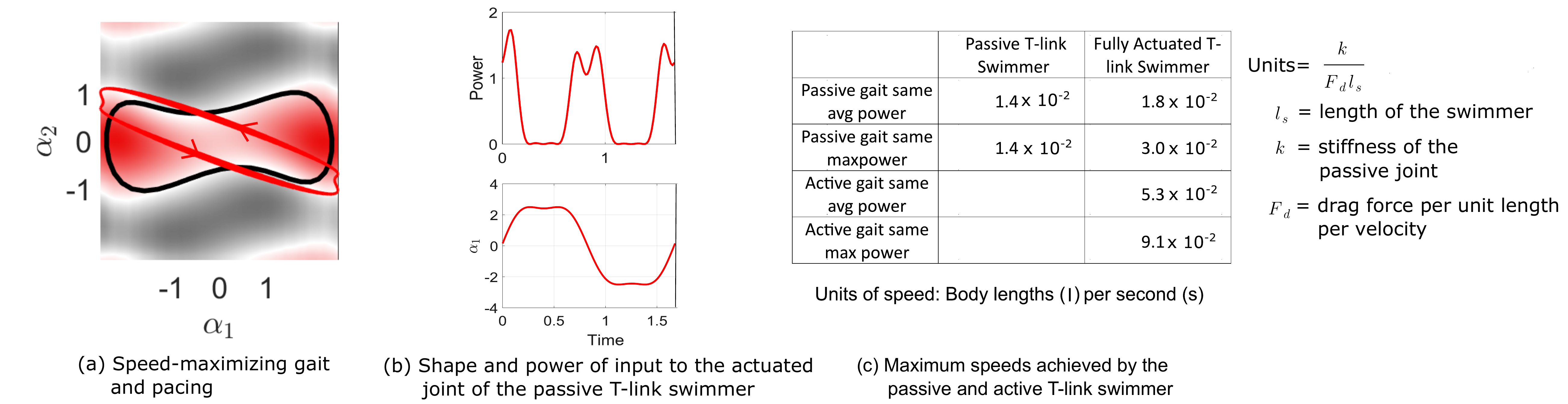}
\caption{Gaits that maximize speed along x-direction for the T-link swimmers. (a) The optimal gait for the passive T-link swimmer (red) and the optimal gait for the fully actuated T-link swimmer (black). The thickness of the line corresponds to the magnitude of power required at different points of the gait. (b) The power required to execute the optimal gait for the passive T-link swimmer and input to the actuated joint over one gait cycle. (c) A comparison table of the speeds achieved by the passive and fully actuated T-link swimmers at different power levels.}
\label{fig:maxspeed}
\end{figure*}

\subsection{Passive Purcell and T-Link swimmers}
We implemented the optimizer described in the \S\ref{sec:Optimizer} in Matlab by providing~\eqref{eq:fullequation1} and~\eqref{eq:fullequation2} as the gradient for the \verb#fmincon# optimizer using the sqp algorithm. The shape of the gait obtained is illustrated in Fig.~\ref{fig:maxspeed3}(a). Fig.~\ref{fig:maxspeed3}(b) shows the power input to the actuated joint over the cycle. Fig.~\ref{fig:maxspeed3}(c) shows a comparison of the speeds achievable by the Passive and fully actuated Purcell swimmers at different power levels. Fig.~\ref{fig:maxspeed} shows the same results for a passive T-link swimmer.

The transfer function relating the response of the passive joint to oscillations of the input joint is given by~\eqref{eq:transfidk}, \(H(s)= \displaystyle {s \over C_{eq} s + K_{eq}}\). Therefore a change in the value of the spring stiffness does not affect the fundamental shape and nature of the response shown by the Bode plot in Fig.~\ref{fig:bode}, but it does shift the entire bode plot to the left or right along the frequency axis. Thus an increase in spring stiffness shifts the Bode plot to the right, which results in the shape of the speed maximizing gait remaining the same, but the time period required to complete the gait decreases, leading to faster speeds.

\section{Energy-Efficient Gaits}
\label{sec:energyopt}

In this section, we describe the gradient calculations involved in identifying the gait that maximizes the efficiency of the swimmers. The objective function we set out to maximize is
\begin{equation}
\eta={g_{\phi} \over E},
\end{equation}
where $g_{\phi}$ is the displacement produced on executing the gait $\phi$ and $E$ is the total energy expended by the robot executing the gait, i.e.
\begin{equation}
E=P_{avg}T
\end{equation}

As explained in~\S\ref{sec:optimalgaits}.B, we parametrize the actuated shape variable using a low-order Fourier series, $p_f$, and obtain the Fourier series parametrization of the resulting motion of the passive joint using the dynamics of the passive elastic joint presented in~\S\ref{sec:frequency}. With these Fourier-series parameters, we can obtain a sequence of waypoints $p_i$ equally spaced in time, that explicitly define the location of the discretization points in the shape space. These waypoints form a direct-transcription parametrization of the gait $p$, obtained from the Fourier series parametrization $p_f$. In this section, we present the calculation of the gradient of efficiency at each of these points with respect to $p$. We then project these gradients onto the Fourier series basis as explained in~\S\ref{sec:fourierpara} to obtain the efficiency maximizing gaits using gradient descent.

The maximum-efficiency cycle must satisfy the condition that the gradient of efficiency with respect to the parameters, $\varparam$ and $T$ is zero, i.e.
\begin{eqnarray}
{\partial \over \partial p}\Big({g_{\phi} \over E}\big)&=&{1 \over E}{\partial g_{\phi} \over \partial p}-{g_{\phi} \over E^2}{\partial E \over \partial p} \nonumber \\
&=&0 \label{eq:energyopt1}
\end{eqnarray}
and
\begin{eqnarray}
{\partial \over \partial T}\Big({g_{\phi} \over E}\big)&=&{1 \over E}{\partial g_{\phi} \over \partial T}-{g_{\phi} \over E^2}{\partial E \over \partial T}\nonumber \\
&=& 0
\end{eqnarray}

Once we have the gradient of efficiency with respect to a direct transcription parametrization $p$ obtained from a Fourier series parametrization $p_f$, we follow the process outlined in~\S\ref{sec:fourierpara} to obtain the gradient of efficiency with respect to the Fourier series parametrization, ${\partial \over \partial \varparam_f}\Big({\gaitdisp \over E}\Big)$.

Thus for suitable seed values of $p_{f0}$ and $T_0$, the maximum efficiency gaits can be obtained by finding the equilibrium of the dynamical system,

\begin{eqnarray}
\dot{p}_f&=& {\partial \over \partial p_f}\Big({g_{\phi} \over E}\Big)\label{eq:effshape}\\
\dot{T}&=&{1 \over E}{\partial g_{\phi} \over \partial T}-{g_{\phi} \over E^2}{\partial E \over \partial T} \label{eq:efftime}
\end{eqnarray}
Thus the process of finding the most efficient gait is the result of a unified process where~\eqref{eq:effshape} is the equation that helps find the shape of the input to the actuated joint and~\eqref{eq:efftime} is the equation that helps find the optimal pacing of the input. Note that as in~\S\ref{sec:Optimizer}, the two equations do not operate independently, and the gradient of shape depends on the time period $T$, and the gradient of the time period depends on the shape of the gait.

\begin{figure*}%
\centering
\includegraphics[width=\textwidth]{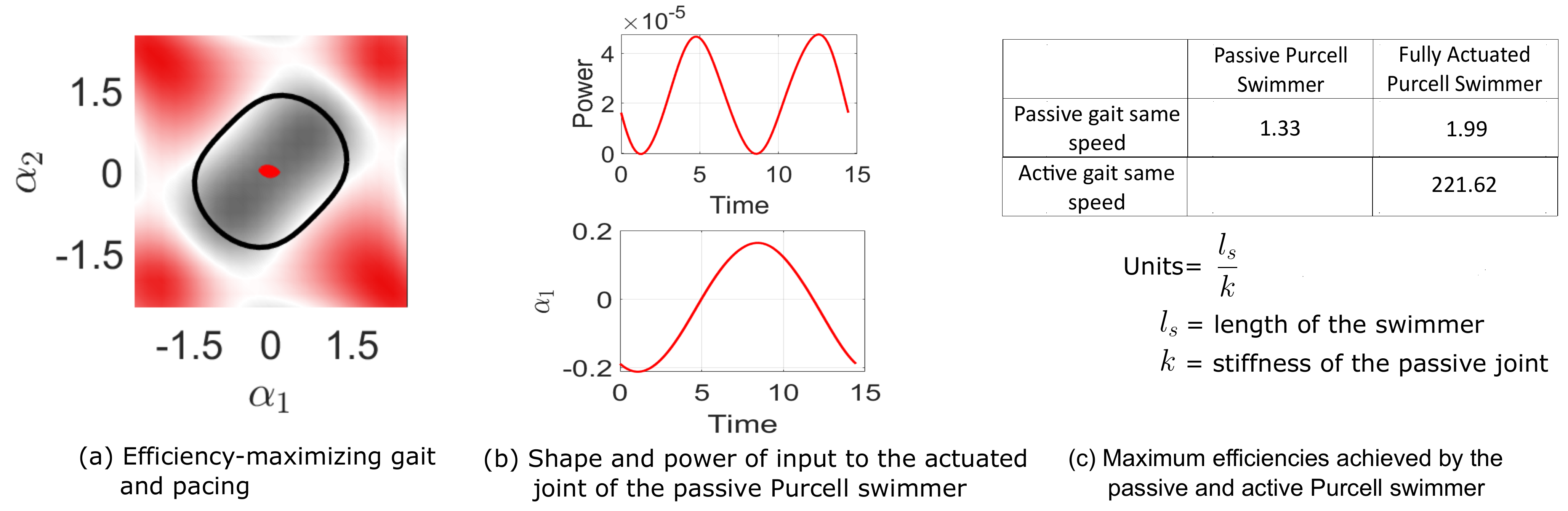}
\caption{Gaits that maximize efficiency along x-direction for the Purcell swimmers. (a) The optimal gait for the passive Purcell swimmer (red) and the optimal gait for the fully actuated Purcell swimmer (black). The thickness of the line shows the magnitude of power required at different points of the gait. (b) The power required to execute the optimal gait for the passive Purcell swimmer and input to the actuated joint over one gait cycle. (c) A comparison table of the efficiencies achieved by the passive and fully actuated Purcell swimmers moving forward at the same speed as the passive swimmer. Note that for the passive Purcell swimmer, the optimizer stops because reducing the frequency further or making the gait smaller does not provide any meaningful increase in efficiency. This observation is line with the results in~\cite{passov2012dynamics}.}
\label{fig:maxeff3}
\end{figure*}

\begin{figure*}%
\centering
\includegraphics[width=\textwidth]{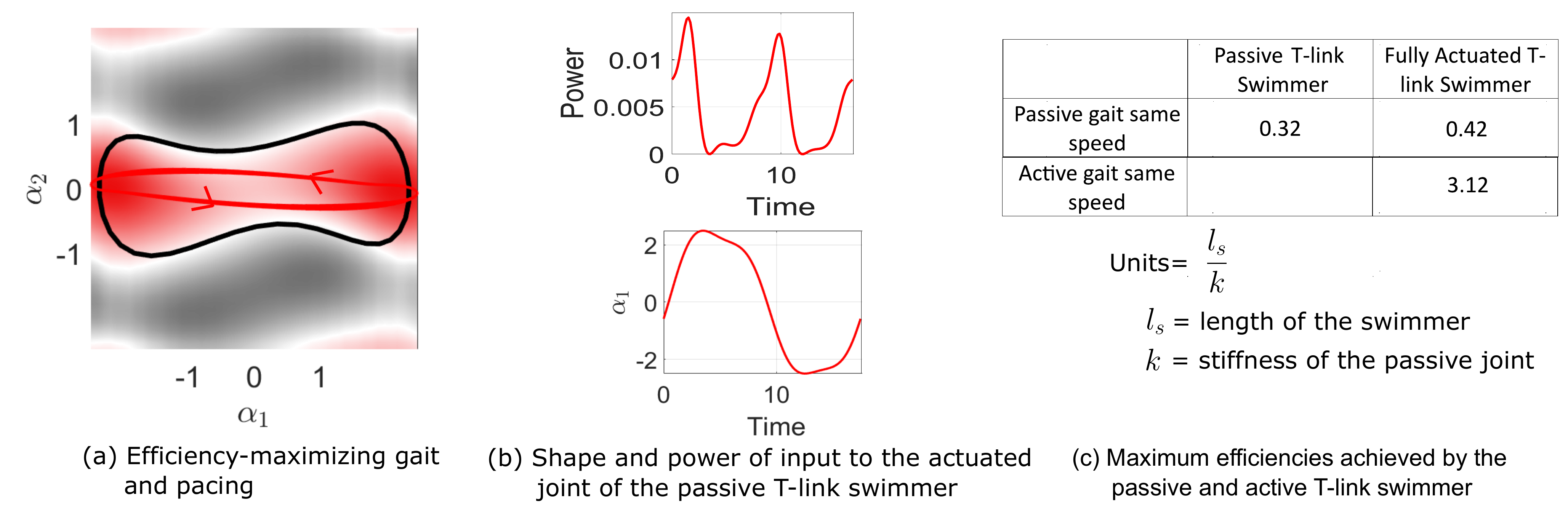}
\caption{Gaits that maximize efficiency along x-direction for the T-link swimmers. (a) The optimal gait for the passive T-link swimmer (red) and the optimal gait for the fully actuated T-link swimmer (black). The thickness of the line corresponds to the magnitude of power required at different points of the gait. The maximum efficiency gait of the passive T-link swimmer is much larger compared to the maximum efficiency gait of the passive Purcell swimmer because a larger gait helps exploit the presence of two peaks in the constraint curvature function of the T-link swimmer. (b) The power required to execute the optimal gait for the passive T-link swimmer and input to the actuated joint over one gait cycle. (c) A comparison table of the efficiencies achieved by the passive and fully actuated T-link swimmers moving forward at the same speed as the passive swimmer.}
\label{fig:maxeffT}
\end{figure*}

\subsection{Shape gradient of the optimal input to the actuated joint}
The shape of the optimal input to the actuated joint is affected by two gradients, ${\partial g_{\phi} \over \partial p}$and ${\partial E \over \partial p}$~\eqref{eq:effshape}. The details of how ${\partial g_{\phi} \over \partial p}$ is calculated are explained in \S\ref{sec:Optimizer}.A. Whereas ${\partial g_{\phi} \over \partial p}$ pushes the gait towards maximum displacement cycles, $E$ is a measure of the cost required to execute the gait and ${\partial E \over \partial p}$ pushes the gait towards low cost shapes. At the most efficient gait, these two opposing gradients cancel each other out, and we get an equilibrium for the gait optimization process.

Over a gait cycle, no energy is stored in the spring. Hence we can calculate the energy expended, $P$, while executing a gait by integrating the power flow through the actuated joint ($\alpha_1$).
\begin{eqnarray}
E&=&\int_0^T\dot{\alpha}_1(t)^T\tau_1 dt \\
&=& \int_0^T\dot{\alpha}_1(t)^T\metric_1(t)\dot{\alpha}(t) dt
\end{eqnarray}
where $\metric_1(t)$ is the first row of the power metric $\metric(t)$.
The gradient of cost with respect to the shape of the gait, ${\partial E \over \partial p}$, is calculated by
\begin{eqnarray}
{\partial E \over \partial p}&=&{\partial  \over \partial p}\int_0^T\dot{\alpha}_1(t)^T\metric_1(t)\dot{\alpha}(t) dt  \\
&=&\int_0^T\Big({\partial \dot{\alpha}_1 \over \partial p}\metric_1\dot{\alpha}+\\
&& \quad \quad \dot{\alpha}_1^T\metric_1{\partial \dot{\alpha} \over \partial p}+\dot{\alpha}_1^T{\partial \metric_1 \over \partial p}\dot{\alpha} \Big) dt
\end{eqnarray}

\subsection{Frequency gradient of the optimal input to the actuated joint}
The equation that governs the optimization process for finding the time period of the most efficient gait is described by~\eqref{eq:efftime}. The term ${\partial g_{\phi} \over \partial T}$ is calculated as described in \S\ref{sec:Optimizer}.B The second gradient in the right hand side of~\eqref{eq:efftime} is calculated as
\begin{eqnarray}
{d E \over d T}={\partial E \over \partial \alpha_1}{\partial \alpha_1 \over \partial T}+{\partial E \over \partial \alpha_2}{\partial \alpha_2 \over \partial T}+{\partial E \over \partial T}.
\end{eqnarray}
Because $\alpha_1$ is the actuated shape variable and the shape of the input actuation is independent of the frequency of actuation, ${\partial \alpha_1 \over \partial T}$ reduces to zero. Therefore the gradient of energy with respect to period reduces to
\begin{equation}
{d E \over d T}={\partial E \over \partial \alpha_2}{\partial \alpha_2 \over \partial T}+{\partial E \over \partial T}.
\end{equation}
The first term accounts for the fact that a change in frequency would change the response of the passive joint $\alpha_2$ resulting in a change in the shape of the gait, and hence a change in the power dissipated while executing the gait. The second term accounts for the fact that even if the shape of the gait remains unchanged, a change in the frequency of input to the actuated joint will change the time required to execute the gait and hence would change the power dissipated while executing the gait.

\subsection{Passive Purcell and T-Link swimmers}
We implemented the optimizer described in the \S\ref{sec:energyopt} in Matlab by providing the gradients of efficiency with respect to shape and time period, calculated using~\eqref{eq:effshape} and~\eqref{eq:efftime} respectively, to the \verb#fmincon# optimizer using the sqp algorithm. The shape of the gait obtained is illustrated in Fig.~\ref{fig:maxeff3}(a) for a Purcell swimmer. Fig.~\ref{fig:maxeff3}(b) shows the power input to the actuated joint over the cycle. Fig.~\ref{fig:maxspeed3}(c) shows a comparison of the efficiencies achievable by the Passive and fully actuated Purcell swimmers when the forward speed for all the systems is fixed to be equal to the forward speed achieved by the passive swimmer when executing its maximum efficiency cycle. Fig.~\ref{fig:maxeffT} shows the same results for a passive T-link swimmer. Note that for the Purcell swimmer, the optimizer stops because reducing the frequency further or making the gait smaller does not provide any meaningful increase in efficiency. This observation is line with the results from~\cite{passov2012dynamics}, where the efficiency was found to asymptotically approach a maximum value as frequency of gait oscillations approached zero. The maximum efficiency gait for the passive T-link swimmer is much larger compared to the maximum efficiency gait of the passive Purcell swimmer because a larger gait helps exploit the presence of two peaks in the constraint curvature function of the T-link swimmer.

As discussed in~\S\ref{sec:Optimizer}.C, change in spring stiffness does not affect the shape and nature of the response of the passive joint, but shifts the bode plot shown in Fig.~\ref{fig:bode} to the left or right along the frequency axis. An increase in spring stiffness shifts the bode plot to the right, which results in the shape of the efficiency maximizing gait cycle remaining the same, but the time taken to execute the gait will decrease. The energy required to execute a gait is inversely proportional to the time taken to execute the gait. Therefore, more energy is required to execute a gait faster. Thus, increasing spring stiffness will result in an overall decrease in the efficiency of swimming.

\section{Comparison with previous work}

\begin{figure*}%
\centering
\includegraphics[width=\textwidth]{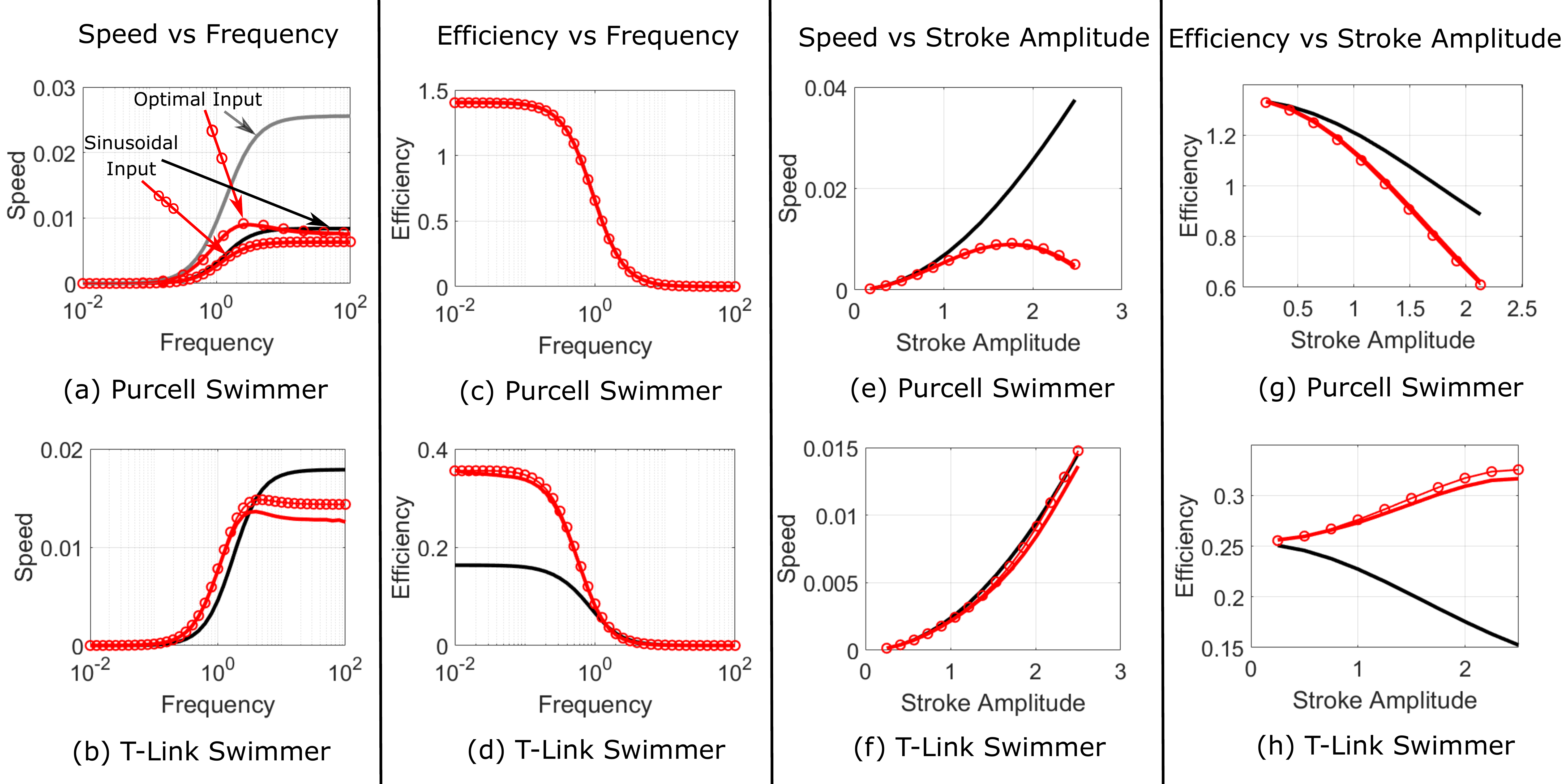}
\caption{In all the subfigures, the solid red lines and the red circles show the speeds and efficiencies predicted by numerical simulation and integral of CCF respectively. The solid black lines show the speeds and efficiencies predicted by the constant-CCF assumption used in~\cite{passov2012dynamics} in the link-attached coordinates. The first column of figures illustrate the speed of the (a) Purcell swimmer as a function of actuation frequency when the input to the controlled joint is a sinusoidal oscillation of unit amplitude and when the input to the controlled joint has the optimal shape and (b) T-link swimmer as a function of actuation frequency when the input to the controlled joint has the optimal shape as obtained in~\S\ref{sec:Optimizer}. In (a), the grey line shows the speed predicted by the constant-CCF assumption for a sinusoidal input. The small amplitude perturbation analysis in~\cite{passov2012dynamics} predicts speed to be a monotonically increasing function of frequency for all inputs. The speed is a monotonically increasing function of frequency for the sinusoidal input. However, it is not monotonically increasing for the optimal input contrary to the prediction from the small amplitude perturbation analysis in~\cite{passov2012dynamics}. The second column of figures illustrate the efficiency of the (c) Purcell swimmer and (d) T-link swimmer as a function of actuation frequency when the input to the controlled joint has the optimal shape as obtained in~\S\ref{sec:energyopt}. Figures (e)-(h) illustrate the speed and efficiency of the Purcell swimmer and T-link swimmer as a function of the magnitude of optimally shaped input actuation obtained from~\S\ref{sec:energyopt}.}
\label{fig:yor_result}
\end{figure*}

Passive swimmers have been previously studied, e.g., in~\cite{passov2012dynamics,krishnamurthy2017schistosoma,montino2015three,jo2016passive,Burton:2010}. Of these works,~\cite{montino2015three,krishnamurthy2017schistosoma} and~\cite{passov2012dynamics} are most relevant to this paper, as they discuss the motion of swimmers with a harmonically-driven active joint and a passive joint. The T-link swimmer used in this paper was introduced in~\cite{krishnamurthy2017schistosoma} as a simplified model to study the swimming dynamics of \textit{Schistosoma mansoni}. The analysis in~\cite{passov2012dynamics} is particularly relevant, as it applies perturbation theory to investigate the motion of the Purcell swimmer.

The approximation used in the perturbation analysis in~\cite{passov2012dynamics} is equivalent to assuming that the displacement produced by executing a gait is equal to the area enclosed by the gait in the shape space multiplied by the constraint curvature value at the center of the gait,
\begin{eqnarray}
g_{\phi}\approx\iint_{\phi_a}D(-\mixedconn)|_\bold{0}=\phi_a \cdot D(-\mixedconn)|_\bold{0}.
\label{eq:yor approximation}
\end{eqnarray}
This approximation has been used in several works from the geometric mechanics community to identify useful shape oscillations that resulted in useful net displacements e.g.,~\cite{Murray:1993,Morgansen:2007}.

Using the approximation in~\eqref{eq:yor approximation}, the authors of~\cite{passov2012dynamics} concluded that for harmonic inputs, the speed of the swimmer monotonically increases with frequency and asymptotically approaches a maximum value as the actuation frequency approaches $\infty$. They similarly concluded that the efficiency of the swimmer asymptotically approaches a maximum value as the actuation frequency approaches zero.

A drawback of this approximation is its the accuracy falls steeply with increasing gait size, as a larger gait would mean larger variations in the value of CCF inside the region bounded by the gait. Our analysis improves on this approximation in three ways: first, our use of a body-averaged frame instead of the link-attached frame in~\cite{passov2012dynamics} ``flattens" some of the nonlinearity in the system, expanding the domain of gait amplitude for which the perturbation analysis gives accurate results. Second, our use of the full integral of constraint curvature over the area enclosed by the gait,
\begin{eqnarray}
g_{\phi}\approx\iint_{\phi_a}D(-\mixedconn), \label{eq:ourapprox}
\end{eqnarray}
absorbs much of the remaining nonlinearity. Third, we use~\eqref{eq:ourapprox} only to calculate gradients, and numerically evaluate the value of $g_{\phi}$ at each step of the gradient descent process described in~\S\ref{sec:Optimizer} and~\S\ref{sec:energyopt} to avoid compounding errors from any residual nonlinearity

Fig.~\ref{fig:yor_result}(a) and Fig.~\ref{fig:yor_result}(b) show how the effect on swimming speed from changing the frequency of the input stroke for Purcell and T-link swimmers respectively. The solid red lines and the red circles show the speeds predicted by numerical simulation and area integral of curvature respectively. The solid black and grey lines show the speeds predicted by the constant-CCF assumption used in~\cite{passov2012dynamics} in the link-attached coordinates for optimally shaped and sinusoidal inputs respectively. The speeds predicted by the constant-CCF assumption are higher than the actual speeds obtained by numerical simulations, and our integral of CCF is a good approximation of the ground-truth simulation.

We can also see from Fig.~\ref{fig:yor_result}(a) that the velocity obtained by the Purcell swimmer does not monotonically increase with frequency for all inputs to the actuated joint. This results in the speed-maximizing gaits found in~\S\ref{sec:Optimizer} having an optimal frequency associated with them, rather than exhibiting a monotonic increase in speed with frequency.

In the case of efficiency-maximizing gaits, for the optimal gait shape obtained in~\S\ref{sec:energyopt}, the efficiency does asymptotically approach a maximum value as shown in Fig.~\ref{fig:yor_result}(c) and Fig.~\ref{fig:yor_result}(d) as the frequency approaches zero. However, the value is different from the maximum efficiency predicted by applying the small perturbation analysis from~\cite{passov2012dynamics} to the T-link swimmer, showing that small perturbation analysis does not completely characterize optimal performance. In the case of the Purcell swimmer, the efficiency-maximizing gait is small enough for the perturbation analysis to yield accurate results.

Figs.~\ref{fig:yor_result}(e)-(h) illustrate how the constant-CCF assumption can introduce errors in identifying optimal actuation shape. Figs.~\ref{fig:yor_result}(e)-(f) illustrate the effect on the swimming speed from changing the size of the input stroke (the reference input is the optimal input obtained in~\S\ref{sec:Optimizer}). Figs.~\ref{fig:yor_result}(g)-(h) illustrate the effect on the swimming efficiency from changing the size of the input stroke (the reference input is the optimal input obtained in~\S\ref{sec:energyopt}). The solid red lines and the red circles show the speeds and efficiencies predicted by numerical simulation and integral of CCF respectively. The solid black lines show the speeds and efficiencies predicted by the constant-CCF assumption used in~\cite{passov2012dynamics} in the link-attached coordinates.

We can see that in the case of the Purcell swimmer, the constant-CCF assumption used in~\cite{passov2012dynamics} incorrectly predicts a monotonic increase in speed with an increase in the amplitude of the input to the actuated joint. In the case of the T-link swimmer, the constant-CCF assumption used in~\cite{passov2012dynamics} incorrectly predicts an increase in efficiency as we shrink the optimal gait. The efficiency would go down if we shrink the optimal gait, because the CCF value for T-link swimmer is higher at the edges than at the center of the shape space. When we shrink the gait, it loses these regions of high value, leading to a decrease in efficiency, which is not captured by the constant-CCF assumption.

The T-link swimmer was first introduced in~\cite{krishnamurthy2017schistosoma}. The analysis in this paper agrees with the most relevant results from~\cite{krishnamurthy2017schistosoma}, which are that when the actuated joint is driven by a simple harmonic input:
\begin{packed_enum}
\ie There exists a linear relationship between the speed-maximizing value of spring stiffness and frequency of actuation.
\ie The average swimming speed increases monotonically as the amplitude of actuation is increased from ${\pi \over 2}$ radians to ${11\pi \over 9}$ radians.
\end{packed_enum}
In~\S\ref{sec:Optimizer}.C, we noted that an increase or decrease in spring stiffness shifts the Bode plot of the response of the passive joint to the right or to the left without changing the shape of the Bode plot resulting in frequency of the speed-maximizing input being  linearly related to the stiffness of the passive joint.

In Fig.~\ref{fig:maxspeed}(a), we can see that the CCF value for T-link swimmer is higher at the edges than at the center of the shape space. Thus an increase in the amplitude of actuation would enclose more of the high value region leading to an increase in speed.

\section{Conclusions}
\label{sec:conclusion}
In this paper, we have identified the geometric structure of optimal gaits for viscous swimmers with passive elastic joints by combining the constraint-curvature analysis in~\cite{ramasamy2019geometry} with frequency-response models for the steady state motion of driven oscillators. We use this structure to identify both speed-maximizing and efficiency-maximizing gaits. The optimal gaits for passive swimmers maximize the CCF integral relative to perimeter and pacing costs, subject to amplitude and phase constraints of a first order system.

As discussed in \S\ref{sec:optimalgaits}, for the fully actuated swimmers, the maximum forward speed achievable is only restricted by the maximum power we are able to supply the joints, but for the swimmers with the passive elastic joint, even with more powerful actuators, there is a theoretical maximum forward speed the system can achieve dictated by the stiffness of the passive joint.

The important factor that makes the performance of the fully actuated swimmers superior to that of the swimmers with the passive elastic joint in terms of energy efficiency is the fact that not only can the fully actuated swimmers execute a much larger set of gaits, they can execute any gait the passive swimmer can execute at a pacing that is just as good or better than the pacing dictated by the dynamics of the passive joint.

This raises the question of what benefits, e.g., simplicity of construction, does having a passive elastic member give to biological organisms that locomote in a low Reynolds number fluid? Most biological organisms have tails that resemble an elastic filament. The propulsive and flexive dynamics of such filaments have been well studied~\cite{wiggins1998flexive,goldstein1995nonlinear,brennen1977fluid,yu2006experimental}. Artificial microscopic swimmers with elastic filaments have been proposed based on this body of work~\cite{dreyfus2005microscopic,edd2003biomimetic}. An interesting line of future work would involve investigating the tradeoff between elastic element inefficiencies and structural complexity of being fully actuated.

This work is the first step towards expanding the applicability of the geometric framework presented in~\cite{ramasamy2016soap,ramasamy2017geometric,ramasamy2019geometry} to systems where underactuated shape parameters play a role in the dynamics of the system. In the case of the passive Purcell swimmer, assuming the torque required to affect a desired shape change did not depend on the current shape of the swimmer did not introduce significant errors in the predicition of the limit cycle corresponding to inputs to the actuated joint as shown in Fig.~\ref{fig:freqfig}. This might not always be the case in all the swimmers we consider. A future line of research would be to improve our frequency domain analysis by using non-linear perturbation theory to obtain more accurate predictions of limit cycles.

Another line of future work would involve studying the shape of optimal gaits in swimmers with more than one passive joint (e.g. a four-link swimmer with two passive joints). The relationship between design choices (e.g. ratio of link lengths, ratio of joint stiffness) and the shape of the optimal gaits would also be an interesting question to answer in systems with more than one passive joint.

\appendix
\section{Comparison with Lighthill Efficiency}\label{appendix:lighthill}
Lighthill's efficiency, defined as the (reciprocal) ratio between the average power consumed by a given stroke and the power that would be required to drag the swimmer at the same average velocity as that produced by the stroke,
\begin{equation}
\eta_{\textsc{lh}}={P_{\textrm{ref}} \over P_{\textrm{avg}}}, \label{eq:lighthill}
\end{equation}
is a commonly-used measure of swimming performance.

Because the drag force acting on the swimmer is proportional to the velocity of its motion, it is readily shown that for a given swimmer, Lighthill's efficiency is proportional to $\tfrac{v_{\textrm{avg}}^2}{P_{\textrm{avg}}}$~\cite{passov2012dynamics}. Ignoring the proportionality factor, we can thus take the Lighthill efficiency as
\begin{equation}
\eta_{\textsc{lh}}={v_{\textrm{avg}}^2 \over P_\textrm{{avg}}},
\end{equation}
where $v_{\textrm{avg}}$ is the average velocity of the swimmer.

In~\cite{ramasamy2019geometry}, we use a geometric measure of efficiency, $\eta_2$, defined in~\eqref{eq:eta2} as the ratio of displacement produced per cycle to the pathlength of the cycle in the shapespace (weighted by the shapespace metric $\metric$),
\begin{equation}
\eta_2={g_{\phi} \over s}
\end{equation}
to identify optimal gaits for fully actuated swimmers. In this appendix, we demonstrate that for fully actuated swimmers the gait that maximizes $\eta_2$, maximizes Lighthill's efficiency and vice versa.

Because $s$ is equal to the time integral of the square root of instantaneous power expended, we can rewrite $\eta_2$ as
\begin{equation}
\eta_2={g_{\phi} \over \int_0^T\sqrt{P(t)} dt}.
\end{equation}
From~\cite{Becker:2003}, we know that the optimal pacing for any gait utilizes a constant power pacing and hence for all $t$,
\begin{equation}
P(t)=P_{\textrm{avg}} \label{eq:appa1}
\end{equation}
Substituting~\eqref{eq:appa1} into the expression for $\eta_2$ provides
\begin{eqnarray}
\eta_2&=&{g_{\phi} \over \sqrt{P_{\textrm{avg}}}T} \\
&=&{v_{\textrm{avg}} \over \sqrt{P_{\textrm{avg}}}} \\
&=& \sqrt{\eta_{\textsc{lh}}}
\end{eqnarray}
Because the square root is a monotonic function, a gait that maximizes our definition of efficiency with respect to path and is executed at constant power pacing also maximizes the Lighthill efficiency optimized over path and pacing, and vice versa.

Note that this conclusion does not hold for the underactuated systems we consider in this paper, for which path and pacing cannot be controlled individually.%

\section{Soap-bubble Gait Optimization}\label{app:soap}

\begin{figure}%
\centering
\includegraphics[width=.45\textwidth]{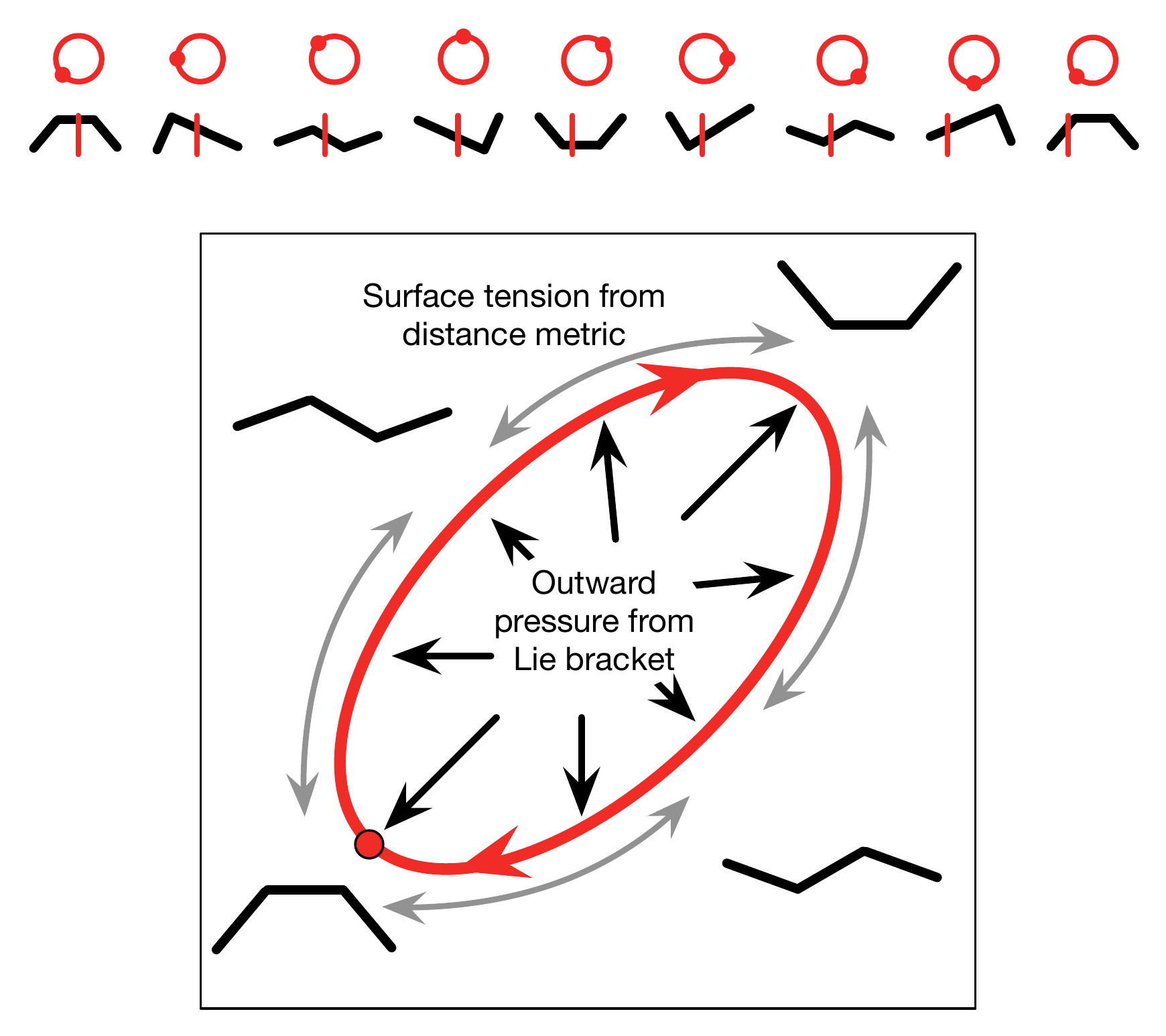}
\caption{Our algorithm in~\cite{ramasamy2019geometry} maximizes gait efficiency in fully actuated swimmers by finding cycles in the space of body shapes that enclose the most \emph{curvature} of the system dynamics (measured via the Lie bracket) while minimizing their cost-to-execute (measured as the metric-weighted lengths of their perimeters). This process is analogous to the process by which air pressure and surface tension combine to produce the shape and size of a soap bubble. Top: The forward progress of a locomoting system as it executes a gait cycle.}
\label{fig:overview}
\end{figure}

\begin{figure*}%
\centering
\includegraphics[width=\textwidth]{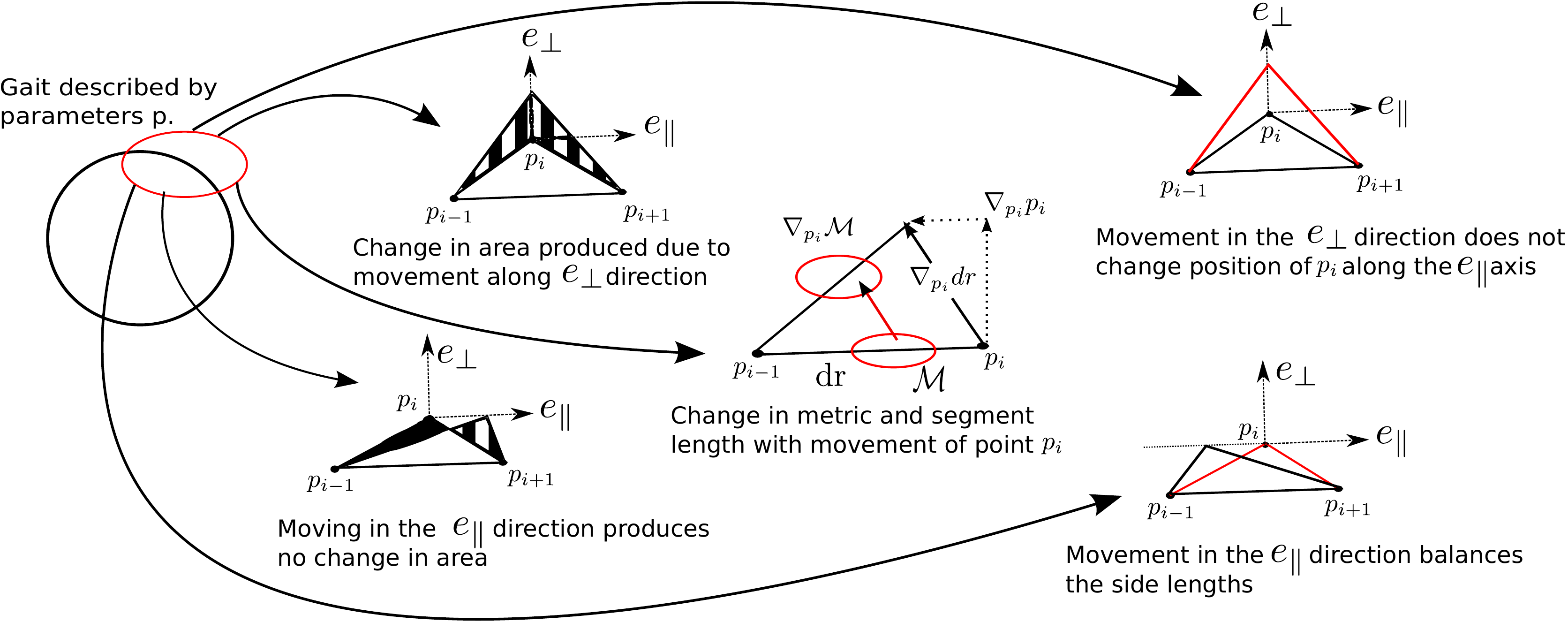}
\caption{Changes in area caused by moving in the two coordinate directions in the local frame. Moving in the tangential direction $\localbasis_{\parallel}$ produces no change in area, as the area of the triangle given by half the product of base length and height remains the same.}
\label{fig:derivation}
\end{figure*}
In this appendix, we present a brief overview of the framework introduced in~\cite{ramasamy2019geometry} to identify optimal gaits for fully-actuated drag-dominated swimmers. In~\S\ref{sec:Optimizer} and~\S\ref{sec:energyopt}, we build on this framework to identify efficiency-maximizing and speed-maximizing gaits for passive swimmers with one passive and one active joint. The framework uses a gradient descent algorithm to identify gaits that maximize efficiency as defined in~\eqref{eq:eta2}.

Given a gait parametrization p, maximum-efficiency cycles satisfy the condition that the gradient of the efficiency ratio is zero,
\beq \label{eq:maxeffcond}
\nabla_{\varparam}\frac{\gaitdisp}{\alnth} = \frac{1}{\alnth} \nabla_{\varparam}\gaitdisp -\frac{\gaitdisp}{\alnth^{2}}\nabla_{\varparam}\alnth= \mathbf{0}.
\eeq

For suitable seed values $\varparam_{0}$, solutions to~\eqref{eq:maxeffcond} can therefore be reached by finding the equilibrium of the dynamical system
\beq \label{eq:gaitequilibria}
\dot{\varparam} = \nabla_{\varparam}{\frac{\gaitdisp}{\alnth}}.
\eeq

The stable equilibria of the right-hand equation in~\eqref{eq:gaitequilibria} are gaits in the same ``image families'' as the system's optimally-efficient gaits (i.e., they follow the same curve as the optimal gait, but not necessarily at the same pacing). To construct the optimal gait, we can either optimize via~\eqref{eq:gaitequilibria} and then choose a constant-metric-speed parameterization, such that the pacing penalty $\intstress$ from~\eqref{eq:intstress} goes to zero, or directly include $\nabla_{\varparam}\intstress$ in our optimizer\footnote{Including $\nabla_{\varparam}\intstress$ in the optimizer works best for parameterizations in which $\nabla_{\varparam}\intstress$ is orthogonal to $\nabla_{\varparam}\frac{\gaitdisp}{\alnth}$, such as waypoint based direct transcriptions. For other parameterizations, e.g., Fourier series, the gradients may not be orthogonal and a two-step procedure of optimizing the image then the pacing will produce better results. For waypoint-based parameterizations, the $\nabla_{\varparam}\intstress$ term has a secondary benefit of helping to stabilize the optimizer by maintaining an even spacing of points, and thereby preventing the formation singularities in the curve)}.

Combining the gradient of the pacing term with the gradient of the image-optimizer places the maximum-efficiency gait as the equilibrium of
\beq \label{eq:fullequation}
\dot{\varparam} = \nabla_{\varparam}\gaitdisp -\frac{\gaitdisp}{\alnth}\nabla_{\varparam}\alnth + \nabla_{\varparam}\intstress
\eeq
(from which we have factored out a coefficient of~$\frac{1}{\alnth}$ from~\eqref{eq:maxeffcond}).

As illustrated in Fig.~\ref{fig:overview}, this differential equation is directly analogous to the equations governing the shape of a soap bubble: $\nabla_{\varparam}\gaitdisp$ takes the Lie bracket as an ``internal pressure'' seeking to expand the gait cycle to fully encircle a sign-definite region, $\nabla_{\varparam}\alnth$ is the ``surface tension'' that constrains the growth of the bubble, and $\nabla_{\varparam}\intstress$ is the ``concentration gradient'' that spreads the soap over the bubble's surface.

For the fully actuated swimmers, we parametrize the gait as a sequence of waypoints $p_i$ such that the gait parameters $p_i$ explicitly define the location of the discretization points. As illustrated in Fig.~\ref{fig:derivation}, each waypoint $p_i$ forms a triangle with its neighboring points and we can define a local tangent direction $e_{\parallel}$ as
\beq
\varparam_{i+1} - \varparam_{i-1} = \ell\, \localbasis_{\parallel}
\eeq
and a local normal direction $e_{\perp}$ orthogonal to $e_{\parallel}$.

We selected this direct-transcription parameterization because it facilitated visualizing the workings of our optimizer (and thus the dynamics governing any other optimization applied to this problem). Additionally, it allowed us to illustrate simultaneous optimization of the gait path and its pacing. We could also parametrize the gait using a Fourier series or Legendre polynomials. In this case, the pacing optimization should be done after the image of the optimal gait has been found because finding an optimal pacing can no longer be formulated as a process orthogonal to the gradient descent process for finding the image of the optimal gait.

\section*{Acknowledgement}

This work was supported by the National Science Foundation, under CMMI grants 1462555 and 1653220. The authors would like to thank Shai Revzen and Brain Bittner for several productive discussions that contributed to this work, and Saad Bhamla for bringing the work in~\cite{krishnamurthy2017schistosoma} to our attention.

\end{document}